\documentclass[fleqn,usenatbib]{mnras}

\usepackage[T1]{fontenc}
\usepackage{ae,aecompl}

\usepackage{graphicx}	
\graphicspath{{imgs/}}
\usepackage{amsmath}	
\usepackage{amssymb}	
\usepackage{xcolor}
\usepackage{color, colortbl}
\usepackage[utf8]{inputenc} 

\newcommand*\diff{\mathop{}\!\mathrm{d}}

\newcommand{\Msun}{\text{M}_{\odot}}	
\newcommand{\HI}{\text{H}_{\text{I}}}	
\newcommand{\fb}{FIREbox$^{\it PF}$}    
\definecolor{mgray}{HTML}{EFEFEF}

\usepackage{newtxtext,newtxmath}

\title[From dark matter to baryons with Deep Learning]{From EMBER to FIRE: predicting high resolution baryon fields from dark matter simulations with Deep Learning}

\author[Bernardini et al.]{M. Bernardini,$^{1}$\thanks{Contact: \href{mailto:mauro.bernardini@uzh.ch}{mauro.bernardini@uzh.ch}},
R. Feldmann,$^{1}$
D. Anglés-Alcázar,$^{2,3}$
M. Boylan-Kolchin,$^{4}$\newauthor
J. Bullock,$^{5}$
L. Mayer,$^{1}$
J. Stadel$^{1}$
\\
$^{1}$Center for Theoretical Astrophysics and Cosmology, Institute for Computational Science, University of Zurich,\\
Winterthurerstrasse 190, CH-8057 Zürich, Switzerland\\
$^{2}$Department of Physics, University of Connecticut, 196 Auditorium Road, U-3046, Storrs, CT 06269-3046, USA\\ 
$^{3}$Center for Computational Astrophysics, Flatiron Institute, 162 5th Ave, New York, NY 10010, USA\\
$^{4}$Department of Astronomy, The University of Texas at Austin, 2515 Speedway, Stop C1400, Austin, TX 78712, USA\\
$^{5}$Department of Physics and Astronomy, 4129 Reines Hall, University of California, Irvine, CA 92697, USA\\
}
\date{Accepted XXX. Received YYY; in original form ZZZ}

\pubyear{2020}

\begin{document}
\label{firstpage}
\pagerange{\pageref{firstpage}--\pageref{lastpage}}
\maketitle

\begin{abstract}
Hydrodynamic simulations provide a powerful, but computationally expensive, approach to study the interplay of dark matter and baryons in cosmological structure formation. Here we introduce the \textbf{EM}ulating \textbf{B}aryonic \textbf{E}n\textbf{R}ichment (EMBER) Deep Learning framework to predict baryon fields based on dark-matter-only simulations thereby reducing computational cost.
EMBER comprises two network architectures, U-Net and Wasserstein Generative Adversarial Networks (WGANs), to predict two-dimensional gas and $\HI$ densities from dark matter fields.
We design the conditional WGANs as stochastic emulators, such that multiple target fields can be sampled from the same dark matter input. For training we combine cosmological volume and zoom-in hydrodynamical simulations from the \textit{Feedback in Realistic Environments} (FIRE) project to represent a large range of scales. Our fiducial WGAN model reproduces the gas and $\HI$ power spectra within 10\% accuracy down to $\sim$10 kpc scales.
Furthermore, we investigate the capability of EMBER to predict high resolution baryon fields from low resolution dark matter inputs through upsampling techniques. As a practical application, we use this methodology to emulate high-resolution $\HI$ maps for a dark matter simulation of a $L=100\,\text{Mpc}\,/h$ comoving cosmological box. The gas content of dark matter haloes and the $\HI$ column density distributions predicted by EMBER agree well with results of large volume cosmological simulations and abundance matching models.
Our method provides a computationally efficient, stochastic emulator for augmenting dark matter only simulations with physically consistent maps of baryon fields.
\end{abstract}



\section{Introduction}\label{sec:introduction}
The fundamental source of cosmological structure formation and its dynamics is the cosmic density field and its non-linear evolution.  The details of the origin and evolution of this structure and its distribution over a variety of scales depends on the physics of the individual matter components -- baryons and dark matter -- and their mutual gravitational interaction. Overdense regions of dark matter, termed dark matter haloes, form the building blocks of large-scale structure as they define the landscape of potential wells in which baryonic matter flows to form individual groups and clusters of galaxies \citep[e.g.][]{Reddick2013, Somerville2015, Guo2010, Wechsler2018, Feldmann2019}. Understanding this coupled evolution in the linear regime is relatively straightforward, but becomes challenging when the evolution transitions into a highly non-linear regime on small scales \citep[][]{Weinberg1972}. Accurately modelling the non-linear interactions on those scales is challenging as they are often intractable for purely analytical approaches. 

\begin{figure*}
    \includegraphics[width=\textwidth]{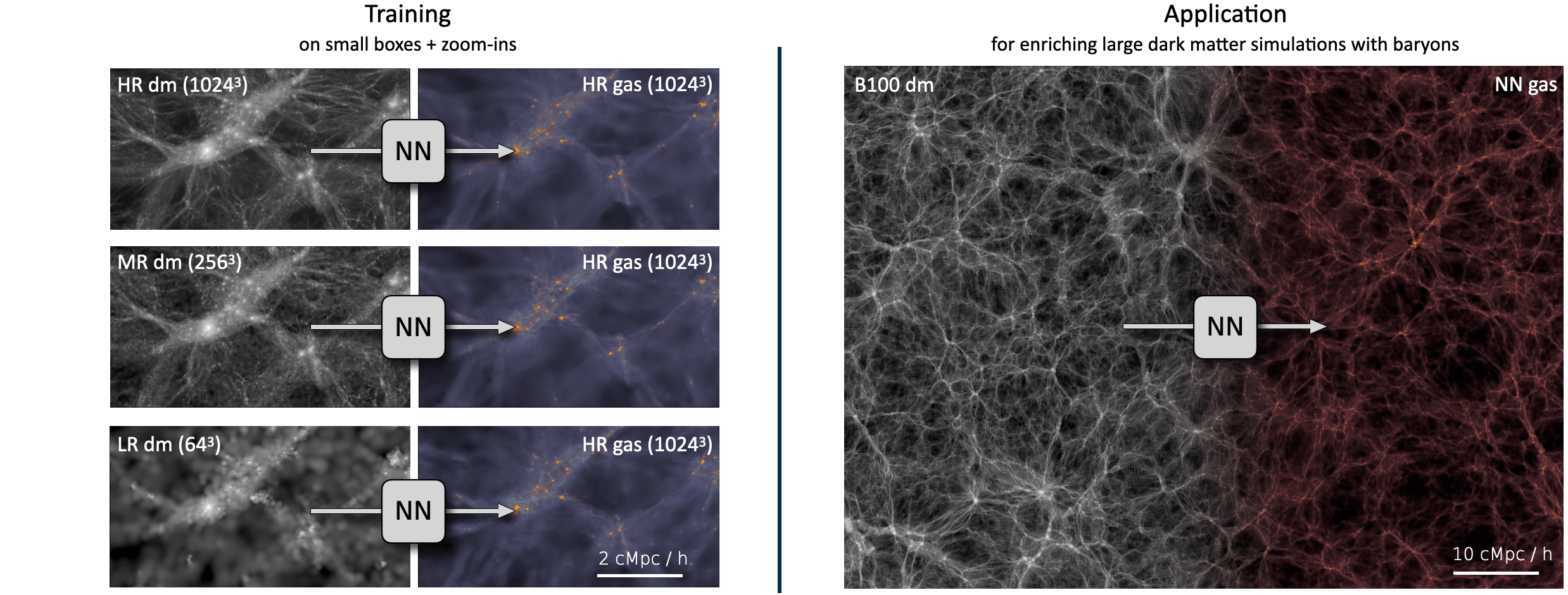}
    \caption{Illustration of our machine learning pipeline. We train neural networks on small cosmological volumes and zoom-in simulations with high resolution to predict baryonic counterparts from dark matter inputs. We investigate the upsampling capabilities of the networks by training individually on different dark matter input resolutions (indicated on the left in the training figure), while the target fields are always fixed to the highest resolution (see section \ref{sec:network_mappings} for details). As indicated on the right, the trained neural networks can then be applied to large dark matter only simulations (e.g. the 100 $h^{-1}$Mpc box used in this work) to enrich them with the specified baryon fields at low computational cost.}
    \label{fig:overview}
\end{figure*}

For this reason, various statistical and semi-analytical models have been proposed to quickly augment dark matter only simulations with information about baryonic components \citep[][]{Somerville1999, Kravtsov2004, Behroozi2013}. Halo Occupation Distribution (HOD) models and abundance matching approaches construct a mapping between the masses of dark matter haloes to the properties of the baryons residing inside them \citep[][]{Peacock2000, Kravtsov2004, Schneider2019}. The mapping is thus governed primarily by the halo mass, neglecting any environmental information such as e.g. the clustering of structure. Semi-analytical models (SAMs) describe the baryonic evolution in more detail \citep[][]{Croton2007, Benson2012}. They use pre-calculated merger trees of dark matter haloes and model the key processes of galaxy formation by a set of coupled differential equations \citep[e.g.][]{Cole2000, Somerville1999, Cora2018}. They are able to take dynamical information into account and can model different gas phases and their interactions. However, SAMs generally do not follow the dynamical interaction of dark matter and baryons and require extensive parameter calibrations \citep[see e.g.][for a detailed comparison of different models]{Knebe2015, Chuang2015}.

To date, numerical hydrodynamical simulations offer the most principled approach to model and study in depth the intrinsic physical properties of systems comprised of dark matter and baryons \citep[e.g.][]{Bryan1998, Springel2003, Springel2005, Keres2005, Vogelsberger2012, Hopkins2014, Wetzel2015, Feldmann2015, Feldmann2016, Wetzel2016, Feldmann2017, Pillepich2017, Hopkins2018, Dave2019}. 
Simulations that consider only the physics of gravity are straightforward; it is baryonic physics, and its backreaction on the dark matter distribution that presents the most substantial challenge at present.
The brute-force computation offers a better understanding of the dark matter and gas dynamics compared to SAMs \citep{Hirschmann2011}.
However, their computational cost, being the main limiting factor, currently prohibits simulations of very large volumes with arbitrarily high resolution \citep[][]{Schaye2010, Schaye2015, Vogelsberger2014, Khandai2015, Feng2016, Dave2016, Nelson2017, Nelson2019}.
The trade-off between simulated box size and particle mass resolution is important, since it limits the range of scales a single simulation can cover \citep[e.g.][]{Katz1993, Knebe2003, Sirko2005, Romeo2008}. Cosmological zoom-in simulations try to mitigate this problem by preselecting a collapse region, which is then enriched in resolution \citep[e.g.][]{Bertschinger2001, Naab2009, Feldmann2011, Hahn2011, Hopkins2014, Angles2014, Onorbe2014}. In this way, very high resolution simulations of individual haloes of different masses are possible, but the technique still suffers from large computational costs and data storage. The major aim of the present work lies in exploring a methodology based on Deep Learning models to overcome this numerical trade-off by enriching cosmological simulations of dark matter with high resolution baryonic information at much reduced computational cost.\\

Feedback processes (e.g. stellar and AGN feedback) regulate star formation by expelling gas back into the surroundings of galaxies \citep[][]{Alcazar2017a, Hopkins2018, Li2018, Biernacki2018, Valentini2019}. As a result, the complicated phase-space and temperature distribution of gas around galaxies inherently contains information about the feedback physics \citep[][]{Barnes2018, Chabanier2020}. Thus, studying absorption signatures of neutral hydrogen and metal lines in the absorption spectra of background quasars is important to reveal the major physical processes that drive galaxy formation.

Galaxies accrete large quantities of fresh metal-poor gas from the intergalactic medium (IGM) to form new stars. This cosmological gas supply has a strong dependence on redshift and halo mass. Gas in massive haloes is shock heated and requires a long time to cool and settle into the galaxy disk whereas cold gas streams can reach the disk directly in less massive haloes \citep[e.g.][]{Keres2005, Dekel2006, Brooks2009, Faucher2011, Woods2014, Ho2019, Stern2020}. This connection between galaxies and gas reservoirs within their parent halo is therefore an important aspect in galaxy formation models.

Modelling the evolution of galaxies requires to understand the evolution of the two main baryonic constituents, stars and gas. Both simulations and observations have led to progress in understanding the evolution of stellar properties such as e.g. the star formation rate (SFR) and the main sequence over cosmic times \citep[][]{Karim2011, Guglielmo2015, Hwang2019, Tacconi2020, Feldmann2020}. The understanding of the dense molecular phase of the interstellar medium (ISM) has also improved through observational surveys of H$_2$ abundances with CO tracing techniques \citep[e.g.][]{Bolatto2013, Tacconi2018, Pavesi2018, Decarli2019}. 

In contrast, much less is known about atomic hydrogen ($\HI$), especially at intermediate to high redshifts. Simulations predict that a significant fraction of the accreted gas in haloes is relatively cold and thus contains large amounts of atomic hydrogen \citep{Keres2005, Fumagalli2011, Faucher2011b, Fumagalli2013, Nelson2013}. The column densities of $\HI$ typically increase towards galaxy centers, which makes absorbers with high $\HI$ column densities better tracers of the gas in the near vicinity of galaxies \citep{Rahmati2014, Bird2014, Crain2016, Diemer2019, Stern2021}. Unfortunately, due to current observational constraints only galaxies residing in relatively massive haloes ($\geq 10^{12} \Msun$) can be observed, which requires large boxes or many individual zoom-in runs to simulate a statistically sound sample of such systems \citep[][]{Altay2011, FaucherGiguere2015, FaucherGiguere2016, Barnes2018}. 

The $\HI$ distribution in galaxies of the local Universe is measured through observations of emissions in the 21cm line \citep[e.g.][]{Kirby2012, Reeves2015}. This method is currently limited to nearby galaxies. After reionisation ($z \sim 6$) the observation of neutral gas is currently only possible through absorption signatures in the spectra of bright background sources (e.g. quasars) \citep[][]{Altay2011, Morganti2018, Glowacki2019, Weltman2020}. Future 21cm observations with significantly improved sensitivity such as e.g. the Square Kilometer-Array \citep[e.g.][]{Weltman2020} will be able to map the distribution of neutral hydrogen at high redshifts, thus enabling to study the formation processes of stars and galaxies in the young Universe \citep[][]{Mellema2013, Pritchard2015, Koopmans2015}.\\

Machine learning and especially deep learning algorithms have recently become a promising tool to capture and learn high dimensional relations related to physical processes in cosmology \citep{Ntampaka2019, Cohen2020, Villaescusa2020, Villaescusa2020b}. They offer a valuable alternative between the computationally much cheaper but low resolution semi-analytical models and the much more expensive hydrodynamical simulations. A major advantage of machine learning methods is that predictions can be produced on time scales typically much smaller than simulations, mitigating a major bottleneck.

Recently, a wide variety of deep learning algorithms have been deployed to accelerate the generation of cosmic matter fields.
Applications range from generating dark matter density fields of different cosmologies \citep{Perraudin2020, Feder2020}, to super-resolution maps \citep{Ramanah2020, Li2020} to weak lensing convergence maps \citep{Tamosiunas2020}.
Another interesting application is to learn the mapping between two matter components. A prominent example is to link the distribution of dark matter to specific baryonic fields like galaxies \citep{Agarwal2018, Zhang2019, Jo2019, Moster2020}, neutrinos \citep{Giusarma2019} as well as various gas fields \citep{Troester2019, Zamudio2019, Dai2020, Thiele2020, Wadekar2020, Lovell2021, Harrington2021, Prelogovi2021, Sinigaglia2021}.

The advancements in adversarial training of neural networks propose an interesting aspect for modelling physical systems that show stochastic variations. 
The key advantage that Generative Adversarial Networks (GANs) offer lies in the probabilistic nature of their predictions. GANs model the underlying distribution of the data by generating samples according to a high-level metric which is learned in the training process.
This is a key advantage compared to neural networks that are trained on low-level metrics (e.g. mean squared error), since the high-level metric is encoded by an entire network itself. When trained correctly, GANs act as stochastic emulators generating samples that are statistically consistent with the dataset. \\

In this work we explore this methodology by combining small high resolution cosmological boxes with zoom-in simulations.
We show that adversarial learning offers a promising pathway for modelling fully non-linear relations between cosmological matter fields. In particular, we use recent advancements in adversarial training of neural networks to predict high resolution baryonic fields from dark matter inputs. We also investigate the upsampling capabilities of the networks by training on different dark matter input resolutions. The trained neural networks can then be used to predict the baryonic counterparts of large dark matter only simulations, which constitutes the primary advantage of this methodology (we show a summary overview in Fig.~\ref{fig:overview}).

This paper is structured as followed. In section \ref{sec:simulations} we describe the simulations used in this work and explain the pipeline to produce the data samples in section \ref{sec:data}. In section \ref{sec:ember_net} we briefly revisit some key aspects in generative learning as well as recent research developments. We introduce the theoretical aspects of the network type used in this work and discuss its key advantages. We also formulate the mappings that the neural networks learn and give a detailed overview of the architectures and further details regarding the training of the networks. The results are presented and discussed in section \ref{sec:results}. Finally, we propose future applications of our method and conclude in section \ref{sec:conclusion}.

\section{Simulations}\label{sec:simulations}
\begin{table*}
\begin{tabular}{llcccccccccc}
\hline
Simulation & note                                 & $L$                  & $N$                         & $m_{\text{b}}$             & $m_{\text{dm}}$            & $M^{\text{max}}_{\text{vir}}$ & $h$    & $\Omega_{\text{m}}$ & $\Omega_{\text{b}}$ & $\Omega_{\Lambda}$ & $\sigma_8$ \\
                           &                 & ($h^{-1}\text{cMpc}$) &                             & ($h^{-1}\Msun$) & ($h^{-1}\Msun$) & ($h^{-1}\Msun$)    &        &                     &                     &                 &  \\ \hline
\textsc{\fb} & hydro   & 15                   & $ 2 \times 1024^{3}$ & $ 4.23 \times 10^{4}$      & $2.27 \times 10^{5}$       & $2.87 \times 10^{12}$         & 0.6774 & 0.3089              & 0.0486              & 0.6911     &   0.8159    \\
\textsc{FIREbox} & dm-only & 15                   & $1024^{3}$ & 0                          & $2.69 \times 10^{5}$       & $4.53 \times 10^{12}$         & 0.6774 & 0.3089              & 0                   & 0.6911      &   0.8159   \\
\textsc{FIREbox} & dm-only & 15                   & $256^{3}$ & 0                          & $1.72 \times 10^{7}$       & $3.03 \times 10^{12}$         & 0.6774 & 0.3089              & 0                   & 0.6911     & 0.8159       \\
A1 & hydro, zoom        & 100                  & $7.50 \times 10^{7}$         & $ 2.32 \times 10^{4}$       & $1.19 \times 10^{5}$        & $1.98 \times 10^{12}$        & 0.6970  & 0.2821              & 0.0461              & 0.7179   &  0.817 \\
A2 & hydro, zoom         & 100                  & $2.33 \times 10^{8}$         & $ 2.23 \times 10^{4}$       & $1.19 \times 10^{5}$        & $2.56 \times 10^{12}$         & 0.6970  & 0.2821              & 0.0461              & 0.7179    &    0.817    \\
A4 & hydro, zoom        & 100                  & $1.34 \times 10^{8}$         & $ 2.32 \times 10^{4}$       & $1.19 \times 10^{5}$        & $2.13 \times 10^{12}$         & 0.6970  & 0.2821              & 0.0461              & 0.7179      &   0.817   \\
A8 & hydro, zoom         & 100                  & $2.92 \times 10^{8}$         & $ 2.32 \times 10^{4}$       & $1.19 \times 10^{5}$        & $2.56 \times 10^{12}$         & 0.6970  & 0.2821              & 0.0461              & 0.7179     &    0.817   \\
B100 & dm-only & 100 & $1024^{3}$ & 0 & $7.9 \times 10^{7}$ & $4.83 \times 10^{13}$ & 0.6774 & 0.3089 & 0 & 0.6911 & 0.8159 \\
\\
\end{tabular}
\caption{Summary overview of all simulations snapshots that are used to produce the network data. $N$ denotes the total initial number of dark matter and gas particles in the simulation volume. $M^{\text{max}}_{\text{vir}}$ is the virial mass of the largest dark matter halo present in that simulation at $z=2$. \fb (hydro) and the zoom-in simulations are used for training and testing the algorithm. The dark-matter-only {\sc FIREbox} runs are used during the resolution study and B100 is solely used for applying our method to a large scale dark matter only simulation.}
\label{tab:1}
\end{table*}
The simulations used in this work are part of the Feedback in Realistic Environments (\textsc{FIRE}\footnote{See the official \textsc{FIRE} project website: \url{https://fire.northwestern.edu}}) project \citep[][]{Hopkins2014, Hopkins2018}. In the following we give a brief overview of the simulation details and show a summary of the most important parameters in table \ref{tab:1}.

We use the $z=2$ snapshots of the \fb and \textsc{MassiveFIRE} simulations run with the FIRE-2 physics model \citep{Hopkins2018} for creating our data sample. Our simulations are run using {\sc gizmo} \citep{Hopkins2015}\footnote{A public version of \textsc{GIZMO} is available at \url{http://www.tapir.caltech.edu/~phopkins/Site/GIZMO.html}}, a multi-method gravity plus hydrodynamics code.
Initial conditions were generated using the multi-scale initial condition tool \textsc{MUSIC} \citep[][]{Hahn2011} where the random seed is fixed. The simulations are run with Planck 2015 cosmology \citep[][]{Planck2016} where $H_0$= 67.74 km/s/Mpc, $\Omega_M$=$0.3089$, $\Omega_\Lambda$=$0.6911$, $\Omega_b$=$0.0486$, $\sigma_8$=$0.8159$ and $n_s$=$0.9667$. Here we summarize only the most important aspects of the simulations and refer the interested reader to corresponding work for further details.

\fb is a high resolution hydrodynamical cosmological simulation with box length of $15\,h^{-1}\text{cMpc}$ (Feldmann et al. in preparation). The box contains initially $1024^3$ dark matter and $1024^3$ gas particles. Dark matter and baryon masses are $m_{\text{dm}}=2.27\times 10^5 \, h^{-1}\Msun$ and $m_{\text{gas}}=4.23\times 10^4 \, h^{-1}\Msun$. The softening lengths for dark matter particles is $80\,\text{pc}$ and gas particles have a minimum softening length of $1.5\,\text{pc}$.

To augment the number of dark matter haloes in the high mass end, we add four zoom-in simulations (A1, A2, A4, A8) selected from the original \textsc{MassiveFIRE} \citep[][]{Feldmann2016, Feldmann2017} suite and re-simulated with FIRE-2 physics and massive black holes \citep[][]{Alcazar2017}, which have a similar resolution as \fb. The particle masses are $m_{\text{dm}}=1.19\times 10^5 \, h^{-1}\Msun$ and $m_{\text{gas}}=2.23\times 10^4 \, h^{-1}\Msun$. The gravitational softening lengths of dark matter and gas particles are 143 and 9 pc.

For this work we make use of the total gas and neutral hydrogen abundances from the simulations. Briefly, gas cooling follows an implicit algorithm described in \cite{Hopkins2018} that includes various processes (free-free, photo-ionization/recombination, Compton, photoelectric, metal-line, molecular, fine-structure, dust collisional and cosmic ray physics). The relevant metal ionization states are tabulated from CLOUDY simulations \citep[][]{Ferland1998} where the process of self-shielding is accounted for via a local Sobolev/Jeans-length approximation calibrated from radiative transfer experiments \citep[][]{FaucherGiguere2010, Rahmati2013}. We refer to \citep[][]{Hopkins2018} for a full description of the simulated gas physics.

We furthermore use a pure dark-matter-only simulation (B100) with boxsize of $L = (100\,h^{-1}\text{cMpc})^{3}$ and dark matter mass resolution of $m_{\text{dm}}=7.9\times 10^7 \, h^{-1}\Msun$ to demonstrate the general applicability of our proposed machine learning approach.
We provide a summary overview of all simulations in Table \ref{tab:1}. 

\section{Data generation}\label{sec:data}
The task of the neural networks is to predict two-dimensional gas and HI mass maps from two-dimensional dark matter inputs. We use a combination of the \texttt{smooth} and \texttt{tipgrid} algorithms for the pixelization of the two-dimensional input and target fields.\footnote{\url{https://github.com/N-BodyShop/smooth}}
For every particle, \texttt{smooth} computes a smoothing length, which is defined as half of the distance to the $n^{\text{th}}$ neighbor particle. We find that setting $n=80$ works well for our approach. Next, we divide the simulation region into 10 equally spaced slabs for each of the 3 spatial directions. \texttt{tipgrid} then interpolates the particles in the same slab onto a two-dimensional grid by depositing the particles mass with a spherically symmetric kernel according to the smoothing lengths computed beforehand. The grid resolution we chose is $4096^{2}$, such that one pixel resolves roughly $3.6\,h^{-1}\text{ckpc}$. We found this combination to be the best for our application. For training the networks we then dynamically create samples with dimensions of $512^{2}$. Although a higher grid resolution would help in resolving even more small scale information, the input dimensions of our neural networks would increase drastically. The resulting set of 30 maps represents the mass of the deposited matter fields over slabs with depth $1.5\,h^{-1}\text{Mpc}$. Note that the largest haloes in our simulations all have $R_{\text{vir}}\sim 350\,h^{-1}\text{ckpc}$ and therefore the probability of splitting a halo between two slabs is negligible for our approach.

A similar procedure is applied to the zoom-in simulations, but here we crop the innermost region containing 75\% of all high resolution particles. Note that for the zoom-in simulations we manually fix the halo center to be at the center of the slabs, and then generate a total of 10 projections according to varying angles. This guarantees, that we only extract the density field within the high resolution region. 

We use the maps from the $x$ and $y$ direction of \fb (hydro) as our training set and retain the projections along the $z$ direction for testing the algorithm. Similarly, we augment the training set with 2/3 of the projections from the zoom-in simulations. The network training uses tiles from the zoom-in simulations and \fb in the exact same way. The maps from the FIREbox $1024^3$ dm-only simulation simply act as an additional testing set for reasons described in the following paragraph.

For our application it is important to notice that the dark matter field in the dark matter only simulations differs from the one in the corresponding hydrodynamical run. We will refer to these fields as dark matter only (\texttt{dmo}) and dark matter hydro (\texttt{dmh}) from now on. The \texttt{dmo} and \texttt{dmh} fields differ on scales where baryonic processes affect the dark matter distribution, such as e.g. adiabatic contraction resulting in deeper halo potential wells due to the cooling of gas in the halo centers \citep{Blumenthal1986, Jesseit2002, Gnedin2004}. Feedback processes can change the underlying dark matter structure as well \citep[][]{Navarro1996, Governato2012, Onorbe2015, CHan2015, Lazar2020}. Another important effect that is induced by the presence of baryons is that the \texttt{dmh} halo morphology tends to be more spherically symmetric compared to the \texttt{dmo} haloes \citep[see e.g.][]{Tissera1998, Bett2010, Kazantzidis2010, Butsky2016, Chua2019, Cataldi2020}.
However, for our application and training algorithm the largest effect is that the exact location of the dark matter haloes is different in the two dark matter fields. Since the gas concentration is typically highest in the halo centers, the \texttt{dmo} dark matter haloes are offset compared to the gas peaks in the target fields. This factor limits the usefulness of the \texttt{dmo} as the input field in the training process, since the neural networks were not designed to learn this shear effect.
We therefore use the \texttt{dmh} version for the traditional training and testing of the networks and retain the \texttt{dmo} version for pure external testing purposes. We then compare the ability of the trained networks to produce the exact power spectrum of the target fields when predicting from either the \texttt{dmo} or the \texttt{dmh} version, despite being only trained on \texttt{dmh} data.
We artificially downsample the $1024^{3}$ simulation to $256^{3}$, $64^{3}$ and $16^{3}$ by randomly selecting particles and adjusting their masses accordingly. The deposition of the downsampled simulations onto the grid is then exactly the same as in the high resolution case (same pixel resolution of $4096^{2}$). We use the suffix \texttt{ds} to indicate downsampled simulations.

The downsampled dark matter maps represent simulations that are in principle different than a simulation of that native resolution, because some high level modes might survive the downsampling process. To understand the impact of training from downsampled dark matter inputs and then applying the networks to dark matter simulations of that native resolution, we also compare all summary statistics to the \fb runs with native resolution $256^{3}$ respectively.

For $\HI$ we also compute the column density distribution function (CDDF) $f(N_{\text{H}_{\text{I}}})$ for the true and predicted maps. The CDDF is a pixel based quantity often used in observational studies of $\HI$, which is defined such that $f(N_{\text{H}_{\text{I}}})\diff N_{\text{H}_{\text{I}}} \diff X$ is the number of absorbers per unit column density bin and unit absorption length $\diff X$. Following \cite{Rahmati2013} we write the CDDF as
\begin{equation}
    f(N_{\HI}, z) \equiv \frac{\diff^2 n}{\diff N_{\HI} \diff X}
\end{equation}
where $\diff X(z)$ is the absorption distance which is related to the box size $\diff L$ as $\diff X =(H_0/c)(1+z)^{2}\diff L$ (see appendix \ref{app:A} for details).

\section{EMBER}\label{sec:ember_net}
EMBER (\textbf{EM}ulating \textbf{B}aryonic \textbf{E}n\textbf{R}ichment) is a framework of several neural networks that we train to map dark matter to baryons. The task of the neural networks is to predict two-dimensional gas counterparts from two-dimensional dark matter inputs.
The pixel resolution of the maps in our dataset is $\sim$3.6 ckpc$/h$, a length scale where astrophysical processes have a large impact on the exact gas configurations.

To understand the importance of modeling such small cosmological scales, we investigate two different methodologies. First, a purely deterministic approach (U-Net), and second, a probabilistic approach (WGAN) that is able to capture the small scale variations in our dataset. Our trained models can then be used as emulators to enrich dark matter simulations with their corresponding gas fields.

In the following sections we describe the theoretical background as well as the implementation of these two methodologies that comprise the EMBER framework. We first introduce both algorithms on a theoretical level, and then discuss implementation and training aspects in more detail. Note that throughout the following sections we refer to the dark matter input maps as $x$ and the target gas fields as $y$ with their corresponding underlying distributions $p_x$ and $p_y$.

\subsection{U-Net}
For the implementation of the neural networks we make use of the U-Net architecture, which was first introduced by \cite{Ronneberger2015} to solve bio-medical image segmentation tasks but has been shown to perform well in regression scenarios as well \citep[e.g.][]{Thiele2020, Wadekar2020}. Our U-Net architecture is a fully-convolutional autoencoder consisting of two main branches, an encoding and a decoding part. As in the general autoencoder case, the information extraction and compression is realized by convolutional blocks followed by information pooling. We use strided convolutions in our implementation. In order to recover the original input dimensions, the decoding units consist of an initial upsampling layer followed by consecutive convolution operations, which results in higher resolution feature maps. In this process the network extracts advanced features, but loses the information of where those features are located in the image. To this end, \cite{Ronneberger2015} introduced skip connections that copy and concatenate the information from the corresponding encoder level with the up-sampled data from the decoder part. In this manner, the spatial information from the contraction path is directly transferred to the expanding branch without being passed through the bottleneck. The skip connections are then concatenated after the upsampling operation with tensors holding information that emerges from deeper parts of the network.

\begin{figure*}
    \includegraphics[width=\textwidth]{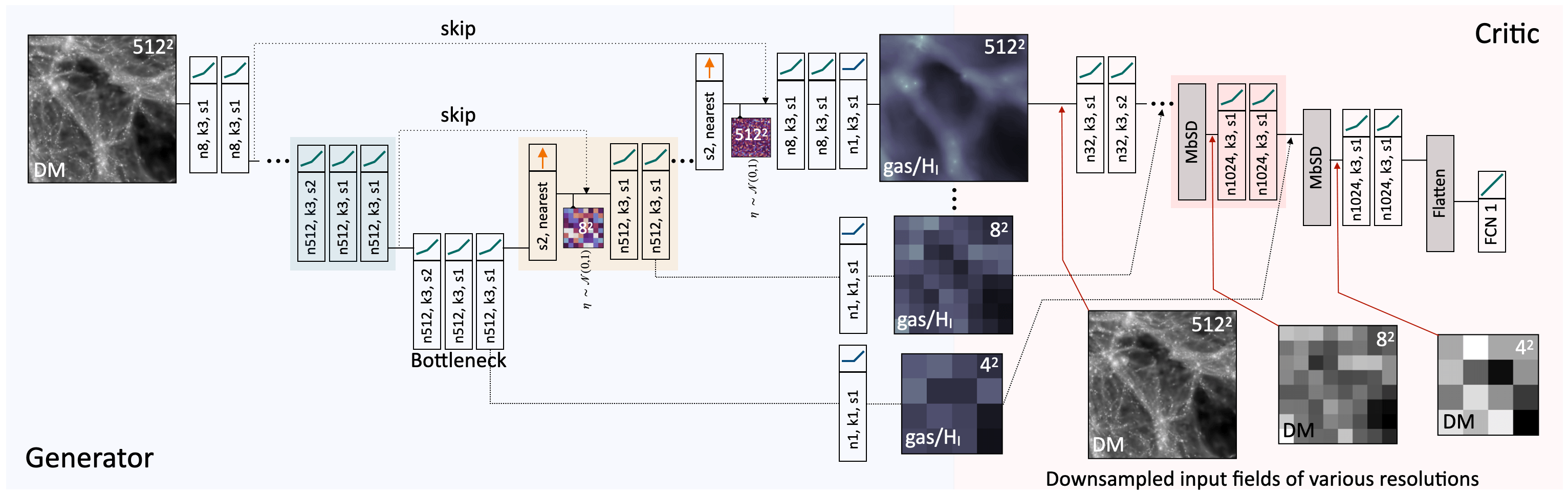}
    \caption{The architecture of the WGAN and U-Net. The generator network shown on the left in light blue is the U-Net with multi-scale outputs whereas the critic is shown on the right in light red. $n, k$ and $s$ denote number of filters, kernel size and stride for each convolution filter in the network, which are all using same padding and are generally activated using LeakyReLUs except for the multi-resolution connections (ReLU) and the final dense layer of the critic (Linear). The green, orange and red shaded blocks of operations are examples of a generator convolution, upsampling and critic convolution block. The noise injection follows the upsampling operation in the decoding part of the WGAN. The artificially downsampled input fields are concatenated in the positions in the critic network as described in section \ref{sec:networks}.
    The pure U-Net also studied in this work  corresponds to the same generator architecture shown above but without any noise inclusion layers and no critic network. The exact same network architecture is used for all trainings and tests of EMBER.}
    \label{fig:network}
\end{figure*}

\subsection{Generative adversarial networks}\label{sec:gans}
Generative adversarial networks (GANs) are a framework first introduced by \cite{Goodfellow2014} where two networks, a generator $G$ and a discriminator $D$, compete in an adversarial game.
$G$ is trained to generate data with a distribution close to the true data distribution $p_y$, where the in- and output are vectors with arbitrary dimensions. $D$ is optimized to distinguish real from fake samples by mapping vectors from the true and generated data domains to [0, 1]. A value of 1 implies that $D$ marks a given sample as real. The training objective of $G$ is to maximize the misclassification of $D$. This setup corresponds to a min-max algorithm where both neural networks try to outperform their corresponding opponent.
To learn an approximation to the true data distribution, one defines an input noise variable $\eta \sim p_\eta$. The generator $G$ learns $p_{g}$, an approximation to $p_y$, through encoding of the latent variable $\eta$. A successfully trained generator can then be used to produce new predictions by sampling the latent space by varying $\eta$.

Conditional GANs (cGANs) incorporate additional information $x \sim p_x$ \citep{Mirza2014}. In practice the conditional information $x$ is an additional feature vector which is used for training both the generator and critic. The min-max game is then expressed in terms of objective functions $\mathcal{L}_D$ for the discriminator and $\mathcal{L}_G$ for the generator \citep{Goodfellow2014} as
\begin{align}
\mathcal{L}_D^{\text{adv}} &= +\mathbb{E}_{\eta} \left[ D(G(\eta|x)|x)) \right] \; - \; \mathbb{E}_{y} \left[ D(y|x) \right ] , \\
\mathcal{L}_G^{\text{adv}} &= -\mathbb{E}_{\eta} \left[ D(G(\eta|x)|x)) \right]
\end{align}
where $y$ and $x$ are samples from the true data distributions $p_y$ and $p_x$. For a successfully trained cGAN the distributions $p_g$ and $p_y$ will be very similar ($p_g$$\simeq$$p_y$). Hence, one obtains a model that is capable of sampling from the joint distribution \citep{Mirza2014, Feder2020}
\begin{equation}\label{eq:prob}
    p(y,x)=p_x(x)p_y(y|x).
\end{equation}
A major challenge while training GANs is the situation when $p_g$ and $p_y$ have little overlap or are disjoint. Prior work \citep[see e.g.][]{Salimans2016, Arjovsky2017a} pointed out that this scenario invokes an inherent instability of GAN training. If the overlap of the two distributions is small, training a discriminator that perfectly separates real and fake samples is relatively easy compared to the task of the generator which exposes a major dilemma. On one hand, the discriminator must perform well enough for the generator to have accurate feedback. On the other hand, the gradients for the generator vanish in case of a perfect discriminator.
Various improvements to the training of GAN models have been presented to artificially increase the amount of overlap between $p_g$ and $p_y$ \citep[see e.g.][]{Jenni2019}, such as Instance Noise \citep{Sonderby2016} and one-sided label smoothing and flipping \citep{Salimans2016}.

\subsection{Wasserstein GANs}
\cite{Arjovsky2017b} introduced WGANs to mitigate the aforementioned problems by using the Wasserstein metric as a measure of similarity between $p_g$ and $p_y$. In this framework, the critic $D$ learns an approximation of the Wasserstein distance. This approach requires $D$ to be Lipschitz continuous over the input domain \citep[see e.g.][]{Arjovsky2017b}. Various approaches have been proposed to satisfy this constraint such as weight-clipping \citep{Arjovsky2017b} or total variational regularization \citep{Zhang2018}. In this work we adopt the gradient penalty scheme introduced by \cite{Gulrajani2017}, where $D$ is constrained as $|\nabla D|=1$, to fulfill Lipschitz continuity. The total critic objective then becomes
\begin{equation}
    \mathcal{L}_{D} = \mathcal{L}_D^{\text{adv}} + \gamma \cdot \mathbb{E}_{\hat{y}} \left[ (||\nabla_{\hat{y}} D(\hat{y}|x) ||_{2} - 1)^{2} \right]
\end{equation}
where $\hat{y}$ are samples drawn from the distribution $p_{\hat{y}}$, which smoothly interpolates $p_g$ and $p_y$, and $\gamma$ is a hyperparameter. We refer the interested reader to \cite{Gulrajani2017} for a detailed description of the gradient-penalty scheme.

\cite{Arjovsky2017b} showed this setup does not suffer from vanishing gradients for the generator in the case of a well-performing critic network. Moreover, in case of optimal training, the critic network is fully converged in every iteration step and propagates the most meaningful gradients back to the generator \citep{Arjovsky2017a, Arjovsky2017b}.

\subsection{On the difficulty of synthesizing high resolution images}
The successful training of networks that are able to synthesize high resolution images of large dimensions is a notoriously difficult task. The major bottleneck lies in the training instability of GANs due to the passage of uninformative gradients from the discriminator to the generator as described above. \cite{Karras2017} proposed a new technique by progressively adding higher resolution layers throughout the training process. Gradually adding higher resolution information is a viable technique to mitigate the missing overlap problem as the network first learns to match the distribution in lower dimensions and slowly migrates towards the full distribution by incorporating higher dimensional information.
In this algorithm, whenever a new layer is added it is gradually faded in such that the training progress of the pretrained part is retained. Even though this technique has been shown to produce state-of-the-art performance, the entire training process remains difficult due to various selections of hyperparameters.

\cite{Karnewar2019} proposed a new method termed MSG-GAN that is based on the idea of gradually matching distributions on multiple resolution scales. Opposed to any progressive adding of layers, in MSG-GAN the entire network composed of generator and critic is trained simultaneously on all resolution levels. The key ingredient lies in multi-scale connections between the generator and critic layers of the same resolution, which allow the gradients to be backpropagated into the various levels directly. \cite{Karnewar2019} find that networks including these connections are less sensitive to the choice of hyperparameters or loss functions. 
The method is robust to different network architectures and drastically helps in the training stability over a wide variety of data sets. This property is especially useful for this work, since we train multiple WGANs with fixed hyperparameters across different datasets.

\subsection{Formulating the network mappings}\label{sec:network_mappings}
We train a collection of neural networks to predict two-dimensional total gas and $\HI$ mass maps from dark matter inputs derived from four different resolution levels: $1024^{3}$ (HR), $256^{3}$ (MR), $64^{3}$ (LR) and $16^{3}$ (VLR). The two-dimensional target fields however are always derived from the $1024^{3}$ (HR) simulation. The data is generated from the \fb and 4 \textsc{MassiveFIRE} zoom-in simulations at fixed redshift $z=2$.
The dark matter input $x$ and the target gas field samples $y$ implicitly define the joint probability distribution $p(y, x)$ in equation \ref{eq:prob}. The marginal distribution $p(x)$ is the probability of a certain dark matter configuration that is computed in the simulation.
In our application, the WGANs learn the distribution $p_g(y|x)$, which is modelled implicitly as it is not constrained with any prior information. Since a single simulation run only probes a sub-collection of all possible samples $x$, the learned distribution by the WGANs $p_g(y|x)$ is an approximated version of the true underlying distribution $p(y|x)$.\\

We train a total of 10 separate neural networks, 8 WGANs and 2 U-Nets, to investigate key questions regarding the mappings depending on the resolution level of the dark matter input. We give a detailed overview of the network architectures in section \ref{sec:networks}.
\begin{itemize}
    \item \textit{HR} $\rightarrow$ \textit{HR}: In the case of HR input, we train 4 networks, two WGANs and two U-Nets (see section \ref{sec:networks} for a detailed comparison). Each pair either predicts the target of total gas or $\HI$. We conduct this comparison between U-Net and WGAN to understand the impact of the generative part when being trained on datasets that exhibit small scale features emerging from fully non-linear baryonic effects in the simulation.\\
    \item \textit{MR} $\rightarrow$ \textit{HR}: For the MR input we train two WGANs, one for total gas and one for $\HI$. This application is different, because a substantial amount of dark matter information is missing in the input fields. The WGAN needs to fill in the information about the missing small scale modes in a physically consistent way. To demonstrate the use-case of this method we apply it to a large dark matter only simulation of boxsize 100 $h^{-1}$Mpc (B100) with the same cosmological parameters and mass resolution as the training set.\\
    \item \textit{LR} / \textit{VLR} $\rightarrow$ \textit{HR}: For the LR/VLR case we repeat the same exercise as in the MR case to understand to what level the WGAN is capable of generating accurate target fields when being presented with dark matter information containing only the very largest modes. The conditional information from $x$ is minimal in this application and the learning of the distributions is mainly driven through the direct feedback from the critic network.
\end{itemize}

\begin{figure*}
    \includegraphics[width=\textwidth]{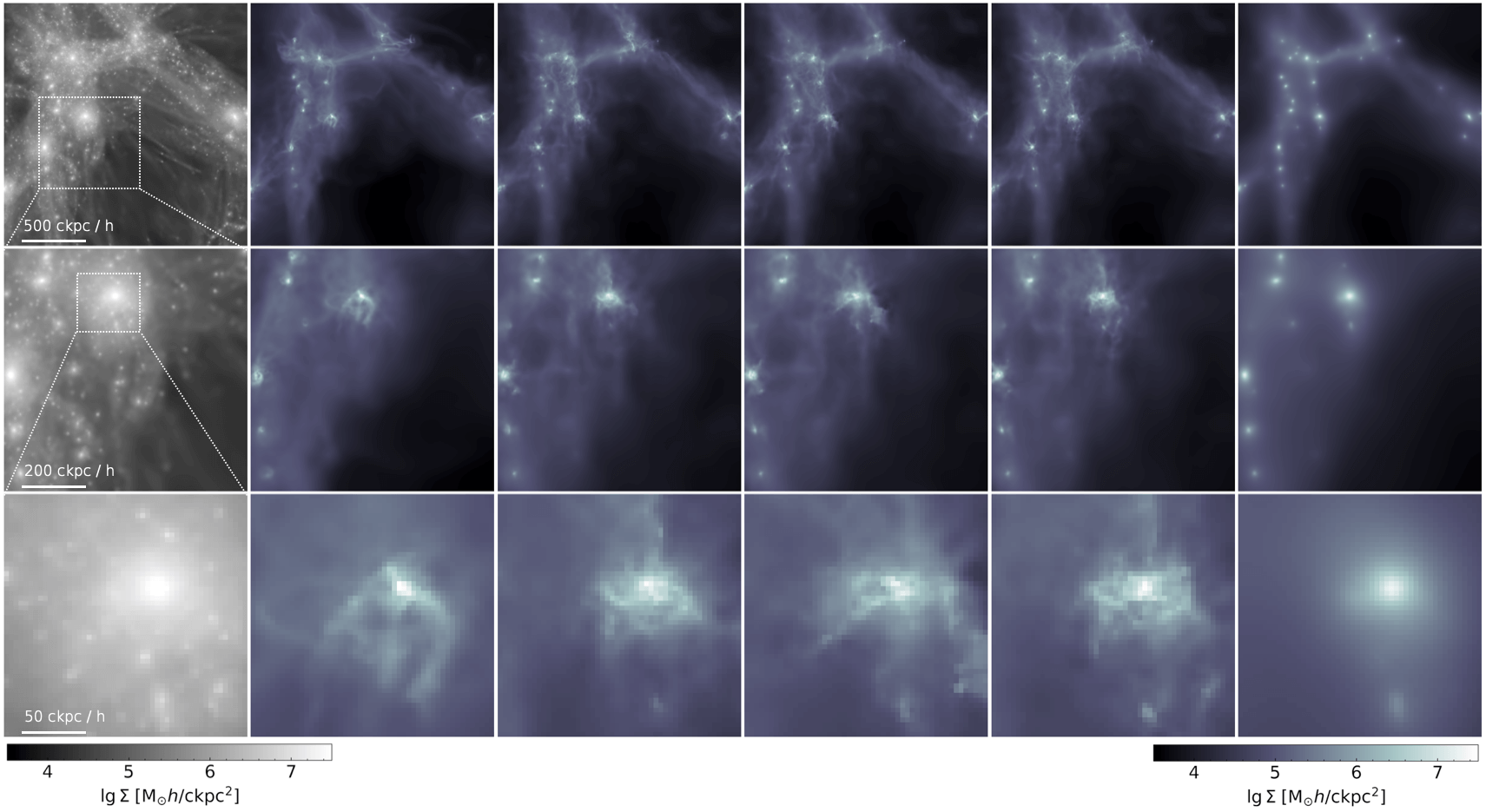}
    \caption{We show a summary grid of the network capabilities regarding visual feature richness for the HR case. The first column is the HR \texttt{dmh} input of the network, whereas the second column shows the corresponding target gas distribution in the \fb simulation. The third, fourth and fifth column are three samples produced by the WGAN and the last column is the U-Net prediction. The tiles in the first row have pixel dimensions of $512^{2}$. We also show zoomed-in regions in the second and third row to highlight the rich features produced by the WGAN on smaller scales.}
    \label{fig:grid_gas}
\end{figure*}
\begin{figure*}
    \includegraphics[width=\textwidth]{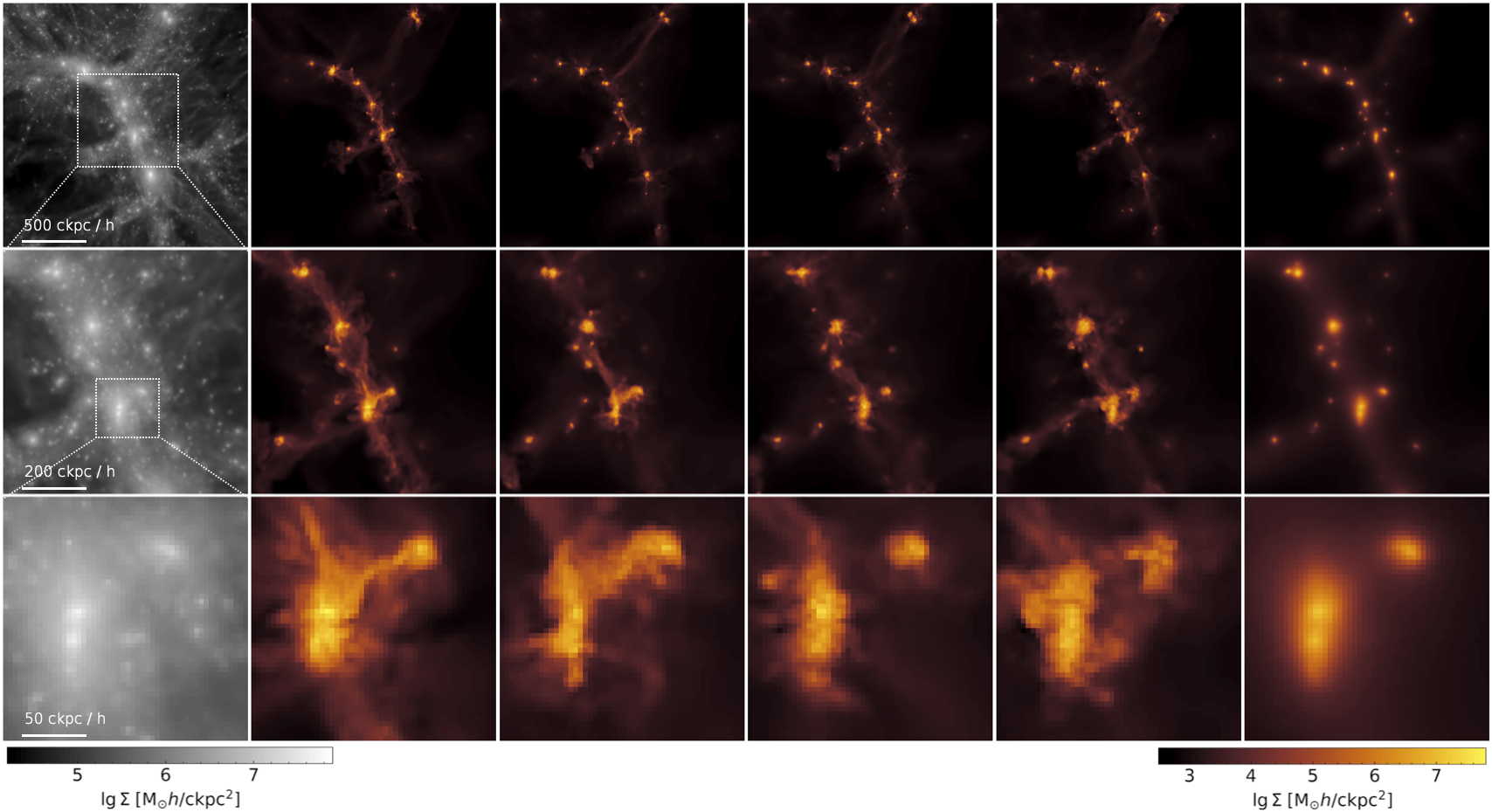}
    \caption{Same as Fig.~\ref{fig:grid_gas} but for the $\HI$ predictions. The log-scaling of the color scheme is the same as in Fig.~\ref{fig:grid_gas} to emphasize the stronger clustering of $\HI$.}
    \label{fig:grid_HI}
\end{figure*}

\subsection{Network Implementation}\label{sec:networks}
We implement the U-Net version in the \texttt{tensorflow} neural network API \citep{tensorflow}.\footnote{Full code can be found at the official \href{https://www.github.com/maurbe/ember}{github} repository.} A detailed schematic of the network is shown in Fig. \ref{fig:network}.
2 The individual convolution blocks are constructed by two convolutions with filter size (3×3) and stride 1. The first network layer has $n_f=8$ convolution filters that are activated by a LeakyReLU. After the first convolution block the data is copied and split along two different paths. The first path is a skip connection that concatenates to the up-flowing data on the reascending network part. Along the second path the data is down-sampled by another convolution layer with $2 \times n_f$ filters, kernel size $(3 \times 3)$ and stride 2 before it flows into the next convolution. All convolution layers are initialized with the glorot-normal initialization scheme \citep[][]{Glorot2010}.

As the data descends the network reaching deeper levels the number of filters increases by a factor of 2 for each subsequent block. Overall there are 8 convolution blocks.
Within deeper layers the network becomes more sensitive to large scale structures in the input image since the strided convolutions reduce the dimensions by a factor of 2 for each additional layer. Since one cell represents a physical size of $\sim$3.6 ckpc$/h$, the downsampling operations increase the receptive field by a total factor of $2^{7}=128$. In the upsampling branch all features scales are used together with the spatial information provided by the skip connections. We use (1x1) convolutions at the end of each upsampling block to obtain the intermediate outputs, which are activated by a native ReLu function as the target fields only contain values greater than 0.\\

Our fiducial WGAN model, is constructed by augmenting the U-Net architecture with techniques from \citep{Karnewar2019, Karras2017, Karras2018}. We mainly follow \cite{Karnewar2019} in our implementation and complement the U-Net with our own critic model which we design to be fully convolutional except for the final dense neuron. Similar to the descending part of the U-Net, the critic is constructed with convolution blocks. A single block contains two convolution layers with filter size $(3\times 3)$ where the second filter has stride 2 for downsampling and for activations we use LeakyReLU. Following \cite{Karnewar2019} we let the critic network extract as much information as possible from all information available at a single iteration step. In order to feed the intermediate scale information from the generator to the critic, we convolve the tensors at the corresponding level with a single $(1\times 1)$ filter \citep{Karnewar2019}. These connections are a vital part as they allow the gradients to penetrate the generator layers on every resolution scale simultaneously. This results in the behaviour observed by \cite{Karnewar2019} where the deepest parts of the generator are optimized first and the synchronization of higher resolution levels follows in a bottom-to-top fashion. The inputs in the first critic layer are the original input field (dark matter) field as well as the target field (total gas or $\HI$). The idea is that the critic network should learn whether or not a certain combination of input and target field is realistic or not on all resolution scales.

The inputs of the lower resolution blocks consist of three fields in total: the information coming from the higher critic block, a downsampled version of the dark matter and the target field itself. We use Minibatch Standard Deviation (MbSD) \citep{Karras2017} on the concatenated tensors coming from the higher level and the downsampled target, but concatenate the downscaled dark matter only after this layer. 
We empirically found that concatenating the dark matter tensor after the standard deviation layer generally results in better performance, presumably because the dark matter information smooths the difference between the outputs of the MbSD layer for a batch of fake and real samples.

The architecture comprised of a U-Net and a critic model defines a prediction pipeline that is deterministic in nature. To promote the generator to a stochastic model, we include an additional noise layer in each upsampling block of the generator to build our fiducial WGAN models. This additional noise then allows the generator to learn a distribution conditioned on the input field and the network is allowed to block any additional noise input, if it does not improve the realism of the generated images. To this end, we follow a similar strategy as \cite{Karras2018}. Instead of concatenating the upflowing tensors with the skip connections directly, the data coming from lower network parts is multiplied with Gaussian noise $\eta \sim \mathcal{N}(0, 1)$ which is controlled by a trainable parameter $\omega$. This parameter can be different for each level as the successful inclusion of noise is in principle resolution dependent. The feature maps $f$ are then modified according to
\begin{equation}
    f' = f + \omega \odot \eta.
\end{equation}
We find that the WGANs make 
use of the additional noise inputs to produce 
small scale details in the generated fields as discussed in section \ref{sec:results}.

\subsection{Training details}
The training of neural networks is simplified when the numerical values of the data are of $\mathcal{O}(1)$. Given that the physical value range of our input and target fields is over nine orders of magnitude, the choice of the data normalization scheme plays a crucial role for our task. We tried many different normalization schemes and found that depending on the cumulative distribution function of the target field the following mixture of a power-law and a log-transform works best,
\begin{equation}
    \Tilde{x} = \frac{1}{k} \log \left[ \left(\frac{x}{x_0} \right)^{q} + 1 \right],
\end{equation}
where $x$ is the field to transform.
The values of the free parameters in this scheme are manually tuned, such that the high end tail of pixel values is retained while keeping intermediate and small pixel values in a reasonable interval as well. 
The exact parameters are given in table \ref{tab:S}.
We found that the appropriate normalization scheme plays a crucial role for achieving accurate predictions for the power-spectrum and bispectrum.

\begin{table}
\begin{tabular}{lllll}
\hline
Field &  & $k$   & $x_0/\Msun$ & $q$   \\ \hline
dmh   &  & 3.0 & $10^{7}$                         & 1.0 \\
dmo   &  & 3.0 & $10^{7}$                         & 1.0 \\
gas   &  & 3.0 & $1.5 \cdot 10^{6}$ & 1.0 \\
$\HI$    &  & 0.9 & $10^{4}$ & 0.2
\end{tabular}
\caption{Summary overview of the manually tuned hyper-parameters in the normalization schemes. The corresponding parameters are the same for normalizing the same field across different resolutions.}
\label{tab:S}
\end{table}

Since the evaluation metrics of the neural network are primarily driven by the high end tail of the pixel distribution, choosing the right loss function for optimizing the networks is a crucial ingredient. In most regression applications, a pure pixel-based loss is enough. We find that adding a loss term measuring the structure similarity of two pictures helps in the training process. We use DSSIM \citep[][]{Wang2004}, a metric that alleviates the pixel-by-pixel comparison as it is evaluated upon sub-windows over the image by comparing various moments (mean, standard deviation and covariance) connected to the image morphology. For maximizing the performance on the high pixel values of density peaks, we combine the DSSIM with a pure pixel-based loss in the form of a mean-square-log error (MSLE). The total perceptual loss on every resolution scale is then aggregated to account for the multi-scale (MS) nature of the mapping,
\begin{equation}
    \mathcal{L}_{\text{p}} =  \alpha \sum_{\text{MS}} \text{DSSIM}(t,p) + \beta \sum_{\text{MS}} \text{MSLE}(t,p),
\end{equation}
where $t$ and $p$ denote true and predicted maps.
Since we only have the generator in the U-Net case, the loss function only contains the perceptual loss. In the WGAN case the perceptual loss is simply added to the adversarial loss for the generator, i.e.
\begin{equation}
    \mathcal{L}_{\text{U-Net}} 
    = \mathcal{L}_\text{p} \quad \text{and} \quad
    \mathcal{L}_G = \mathcal{L}_{\text{p}} + \gamma \mathcal{L}_{G}^{\text{adv}}.
\end{equation}
The prefactors ($\alpha, \beta, \gamma$) account for weighting the different loss contributions (see table \ref{tab:loss_weights}). We note that even though the adversarial loss could in principle take care of any pixel-based loss by simply encoding it, we find that keeping $\mathcal{L}_{\text{p}}$ helps with convergence and stability especially in the beginning parts of training. Once the perceptual loss saturates, the gradient penalty and adversarial losses become the dominant measure of the WGAN performance.

\begin{table}
\begin{tabular}{llll}
           & $\alpha$ & $\beta$ & $\gamma$ \\ \hline
U-Net (gas) & 1                    & 2                   & -  \\
U-Net ($\HI$)  & 1                    & 200                 & - \\
WGAN (gas) & 10                   & 0.4                 & 10  \\
WGAN ($\HI$)  & 5                    & 0.02                   & 50                    
\end{tabular}
\caption{Summary table of the chosen hyperparameters used in the total loss function. Since we keep the network parameters fixed across all training runs, ($\alpha, \beta, \gamma$) represent the important hyperparameters in our approach. For the U-Nets we normalize the losses to the DSSIM ($\alpha$). The $\beta$ parameters vary as well since the PDF of the target fields is different. In the WGAN case, the losses are normalized to the adversarial loss.}
\label{tab:loss_weights}
\end{table}

During training time we dynamically create batches of 8 images with size $512^{2}$ from the training set. The optimizer we use is Adam \citep[][]{Kingma2017} with parameters $\beta_{1}=0$ and $\beta_{2}=0.99$ where the learning rates are reduced in a polynomial fashion. The networks were trained for $3\times 10^5$ iterations, which means that every generator network has seen approximately 2.5 million data samples. The training was performed on a single Tesla V-100 GPU card and took 9 hours in the case of the U-Nets and 96 hours for the WGANs.

During prediction time, we only make use of the trained generator to produce new samples. Since the generator architecture is fully convolutional, the network can make predictions on inputs with arbitrary input sizes.
Predicting all projections (of size 4096$^2$) of a single simulation box takes approximately 60 seconds on the same V-100 GPU card.

\section{Comparison and discussion}\label{sec:results}
In this section we give a detailed overview of various statistics that we use to measure the network capabilities. To evaluate the predictive power of the networks we compare the following summary statistics: total mass, pixel probability density, 2-dimensional power spectrum and bispectrum between the true and predicted mass maps. Furthermore, we conduct a halo-based analysis by comparing true and predicted gas masses inside dark matter haloes. In the $\HI$ case we also compare the column density distribution function (CDDF) $f(N_{\text{H}_{\text{I}}})$ between the true and predicted maps.

We observe that in the final stages of training, the predicted power spectrum can fluctuate especially on scales $\sim10\, \text{ckpc}/h$. We use the power spectrum of the prediction on the training set as the primary metric to determine the best networks. We then identify for each U-Net and WGAN the point to stop training, and take this exact checkpoint for further analysis.

For each of the different input resolutions, we use the WGANs to predict a total of 128 test boxes (from \texttt{dmo ds}, from \texttt{dmh ds} and in the 256 case also from the \texttt{dmo native}) to quantify the amount of internal scatter in the summary statistics. Note that each realization corresponds to a different latent vector $\eta$ as described in section \ref{sec:gans}. The U-Nets predict only one box per input as the mapping is deterministic. The statistics of those predicted maps are always compared with the high resolution gas fields (\textit{HR}, $1024^3$), since we want to quantify to what extent we can use lower resolution N-body simulations to achieve similar accuracy compared to the high resolution results.

\subsection{WGAN vs U-Net}
Fig.~\ref{fig:grid_gas} and \ref{fig:grid_HI} are a visual summary showcasing the network capabilities of generating realistic samples across different scales. The WGANs have learned to encode the additional noise inputs resulting in very diverse small scale structures. The last column shows the U-Net predictions, which resemble a smoothed out version of the WGANs. The absence of a critic network paired with no noise inclusion results in very smooth gas maps that lack the high resolution small scale features. From a conceptual point of view this behaviour is understandable because the U-Net performs a regression of the mean of all possible gas configurations.

\begin{figure*}
    \includegraphics[width=\textwidth]{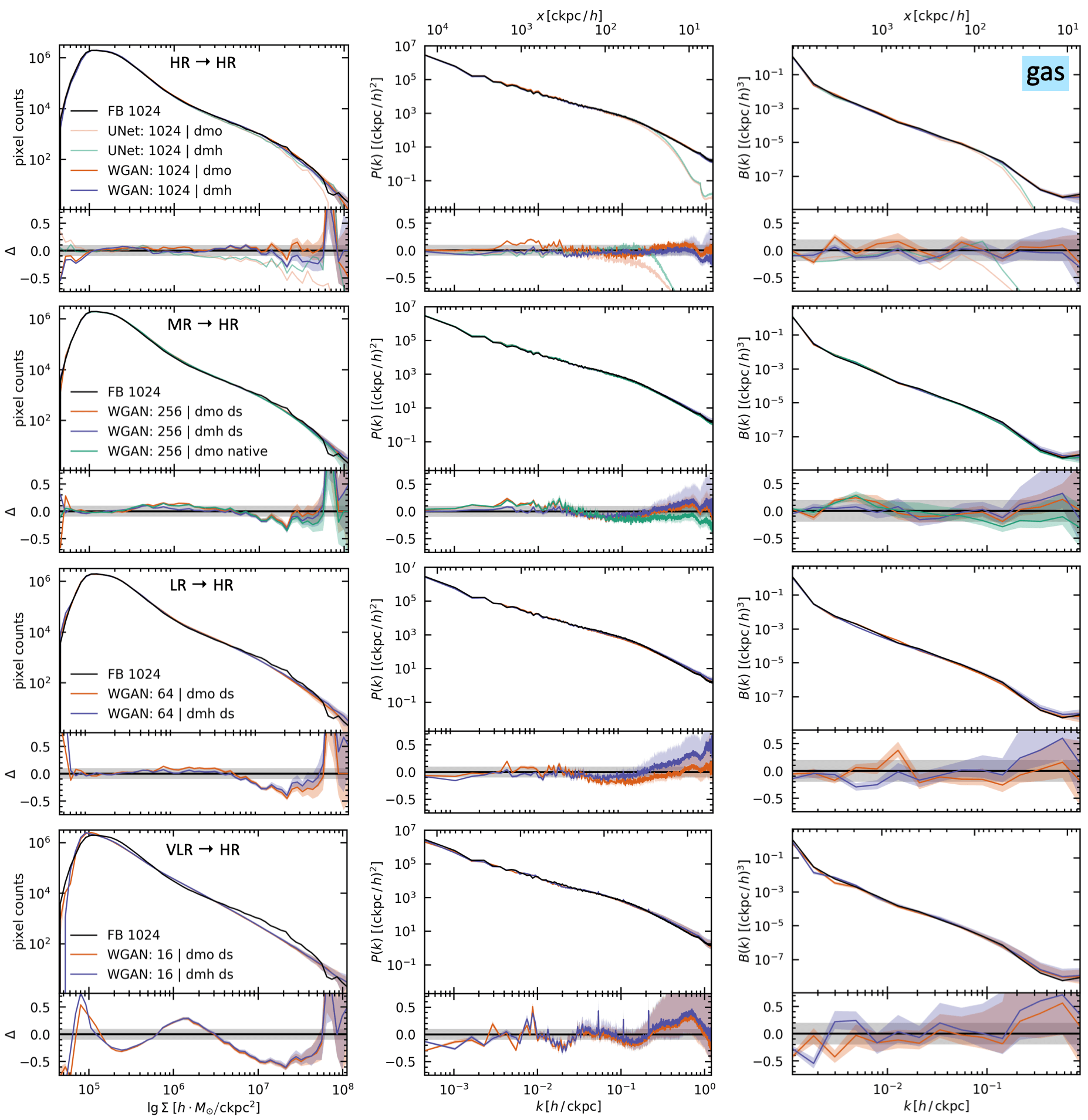}
    \caption{PDF, power spectra and bispectra (from left to right) for the predicted gas projections for WGAN and U-Net when mapping from \texttt{dmh} and \texttt{dmo} of the \fb simulation. The bispectra are shown for the case of equilateral triangles. The four rows represent the statistics for the \textit{HR}, \textit{MR}, \textit{LR} and \textit{VLR} mappings. FB 1024 denotes the \fb \textit{HR} statistics. Note that e.g. (1024 | dmo) indicates that the 2D maps from the $1024^3$ dark-matter only simulation are used as network input. The lower panels show the fractional error when compared to the ground truth. Note that the error is computed on the quantities directly and not on the $\log$ of the quantities. The shaded bands denote the 10\% or 20\% fractional error limit, depending on the statistic that is shown. For the WGANs we plot quantiles (16, 50 and 84) derived from the 128 predicted boxes as a true scatter estimate of the WGAN. All curves show predictions on the test set.}
    \label{fig:gas_summary}
\end{figure*}

\begin{figure*}
    \includegraphics[width=\textwidth]{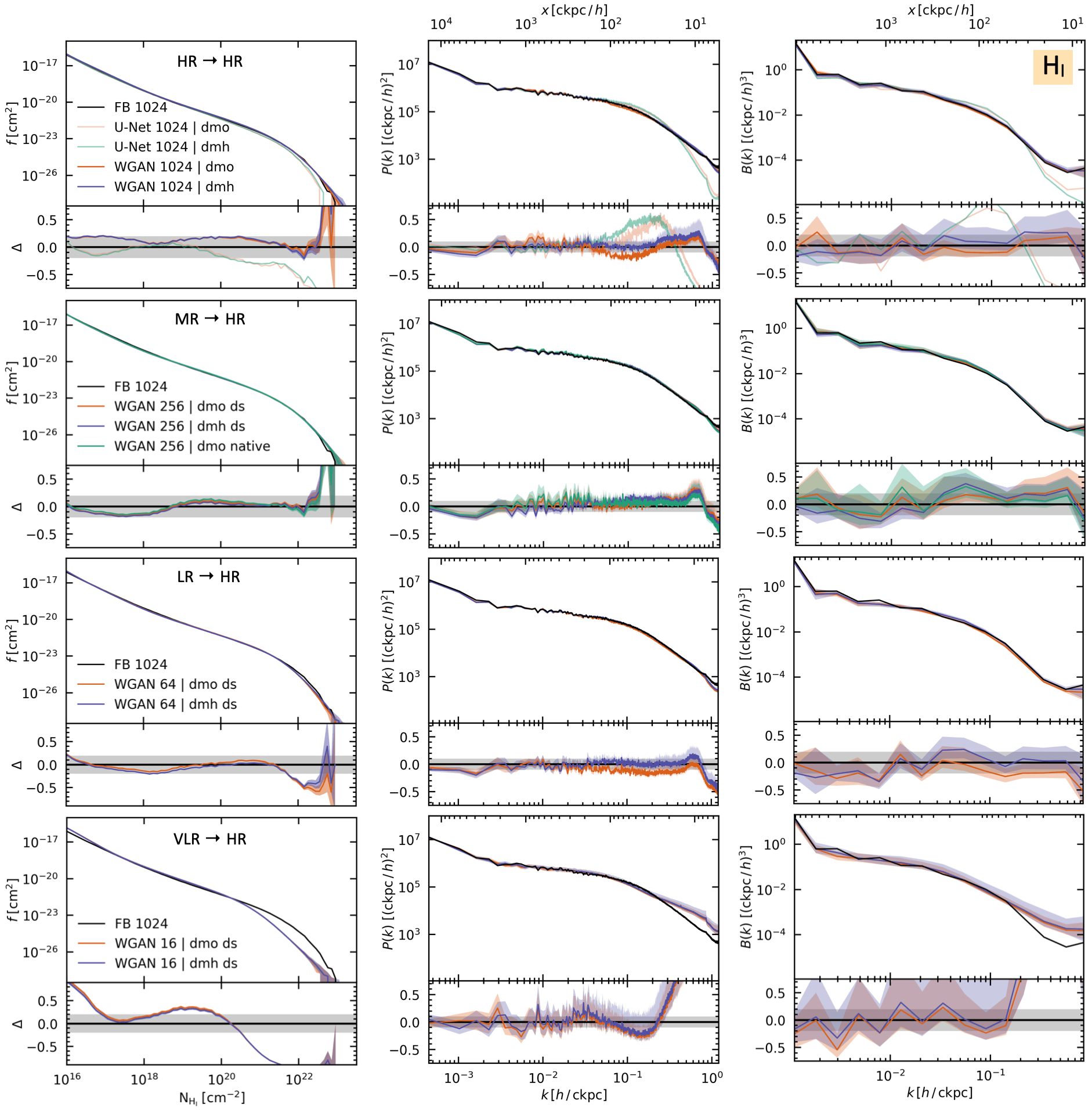}
    \caption{Similar to Fig.~\ref{fig:gas_summary} but showing the column density distribution function, power spectra, and bispectra (from left to right) of the $\HI$ gas component. The shaded bands denote the 10\% or 20\% fractional error limit, depending on the statistic that is shown. As for the gas case we plot quantiles (16, 50 and 84) derived from the 128 predicted boxes as a true scatter estimate of the WGAN. The $\HI$ maps are predicted from the test data sets of the hydrodynamical (\texttt{dmh}) and the dark-matter-only (\texttt{dmo}) \fb simulation.}
    \label{fig:HI_summary}
\end{figure*}

In table \ref{tab:masses} we show the fractional deviation of the total mass in the box between predicted and true gas maps for the eight networks. Generally, the two mappings from either \texttt{dmo} and \texttt{dmh} result in very similar deviations. For the total gas mass, the U-Net and WGANs perform equally well, but the WGANs outperform the U-Nets across all resolutions in the case of the more difficult $\HI$ mapping. In this case we found that the U-Nets perform worse and generally overpredict the mass in the box. In Fig.~\ref{fig:gas_summary} and \ref{fig:HI_summary} we show the predicted PDF, CDDF, power spectra and bispectra for the \textit{HR} case (top row). As expected, the U-Nets fail to reproduce the correct power on small scales whereas the WGAN models show very good agreement for all statistics.

\begin{table}
\begin{tabular}{lllr}
\hline
Network & map                         & input res.                         & $\Delta_{\mathrm{m}}$ [\%] \\ \hline
        &                             &                                    &              \\
\rowcolor{mgray} WGAN    & dmo / dmh $\rightarrow$ gas & $1024^3$ native &       $1.92$       \\
U-Net    & dmo / dmh $\rightarrow$ gas & $1024^3$ native &       -1.13 / -0.56      \\
        &                             &                                    &              \\
WGAN    & dmo $\rightarrow$ gas       & $256^3$ native&       $1.71$      \\
WGAN    & dmo / dmh $\rightarrow$ gas & $256^3$ ds      &       $1.45$ / $1.22$          \\
        &                             &                                    &              \\
WGAN    & dmo / dmh $\rightarrow$ gas & $64^3$ ds       &       $0.86$ / $0.86$        \\
WGAN    & dmo / dmh $\rightarrow$ gas & $16^3$ ds       &       $-8.12$ / $-8.12$        \\ \hline
        &                             &                                    &              \\
\rowcolor{mgray} WGAN    & dmo / dmh $\rightarrow$ $\HI$  & $1024^3$ native &       $5.54$ / $5.54$       \\
U-Net    & dmo / dmh $\rightarrow$ $\HI$  & $1024^3$ native &       -40.84 / -37.33        \\
        &                             &                                    &              \\
WGAN    & dmo $\rightarrow$ $\HI$        & $256^3$  native &       $2.53$        \\
WGAN    & dmo / dmh $\rightarrow$ $\HI$  & $256^3$ ds      &       $4.85$ / $3.9$      \\
        &                             &                                    &              \\
WGAN    & dmo / dmh $\rightarrow$ $\HI$  & $64^3$ ds       &       $-14.34$ / $-14.34$       \\
WGAN    & dmo / dmh $\rightarrow$ $\HI$  & $16^3$ ds       &       $-75.76$ / $-75.76$       \\ \hline
\end{tabular}
\caption{Summary table of the median fractional error (given in \%) of the predicted total mass in the box for the total of eight networks across all tested simulation resolutions and input fields. We denote downsampled dark matter inputs with "ds", whereas otherwise the native resolution is taken as input. All masses are computed and compared to the (ground truth) \fb ($1024^3$) test dataset. The fiducial WGAN models achieve uncertainties of $\sim 2 \%$ for total gas and $\sim 5 \%$ for $\HI$. Generally, predicting the targets from lower resolutions results in larger errors and predicting $\HI$ masses is systematically more difficult than total gas masses.}
\label{tab:masses}
\end{table}


\subsection{WGAN for extreme upsampling}
In the section before we described the performances of the network when trained on  HR dark matter input. In this section we explore the upscaling capabilities of the WGANs when trained on lower resolution dark matter inputs and discuss the impact on the predictive power. 

For the MR case we show the same summary statistics in Fig.~\ref{fig:gas_summary} and \ref{fig:HI_summary}. In general, the mappings between dark matter and gas become more difficult in the case of lower resolution inputs because small scale features are absent in the input fields. Therefore, an accurate encoding of individual features through the noise inputs is necessary to obtain accurate summary statistics. Fig.~\ref{fig:gas_summary} and \ref{fig:HI_summary} show that the WGAN successfully reproduce power spectra within $\sim$ 10\% and bispectra within $\sim$ 20\% down to scales of $\sim$ 10 ckpc/$h$. Interestingly, in the extreme upsampling cases from \textit{LR/VLR} $\rightarrow$ \textit{HR} the WGAN still performs well on the power and bispectra down to $\sim 50 \, \text{ckpc}/h$. However, the pixel based statistics as the PDF and the CDDF show larger deviations, presumably because the power and bispectra are predominantly determined by the high density pixels, whereas the PDF and CDDF depend on the entire pixel range. In particular, our method fails to reproduce any sensible prediction of the CDDF in the extreme upsampling case of $\textit{VLR} \rightarrow \textit{HR}$.

\subsection{Predictive power of EMBER}
In this section we quantify the predictive power of EMBER by exploring how well the networks perform when tested on physically motivated metrics. Furthermore, we apply the WGAN model to a large dark matter only simulation (B100) to extend the metrics and compare them to related work.

\subsubsection*{Halo based analysis} \label{sec:halo_analysis}
The WGAN methodology defines a framework that is completely halo-free. It is therefore interesting to conduct a halo-based analysis despite the model not knowing the notion of a dark matter halo. For this we compare the projected gas masses within one projected virial radius to the corresponding dark matter masses of individual haloes. 

First, we identify dark matter haloes with \texttt{Amiga Halo Finder} \citep[][]{Knollmann2009} \footnote{Code available at: \url{popia.ft.uam.es/AHF/Download.html}.}. We then use the information of the virial radii to construct 2-dimensional masks and compute the included dark matter and gas masses from the projected density fields.
We note that this approach computes the masses over the entire $1.5\,h^{-1}\text{cMpc}$ which would in principle overestimate $\HI$ masses by including gas outside $R_{\text{vir}}$. However, as shown in Feldmann et al. (in prep), at $z=2$ almost all $\HI$ resides inside dark matter haloes and the amount of $\HI$ beyond one virial radius is negligible. We refer to App. \ref{app:B} for a more detailed discussion.

\begin{figure*}
    \includegraphics[width=\textwidth]{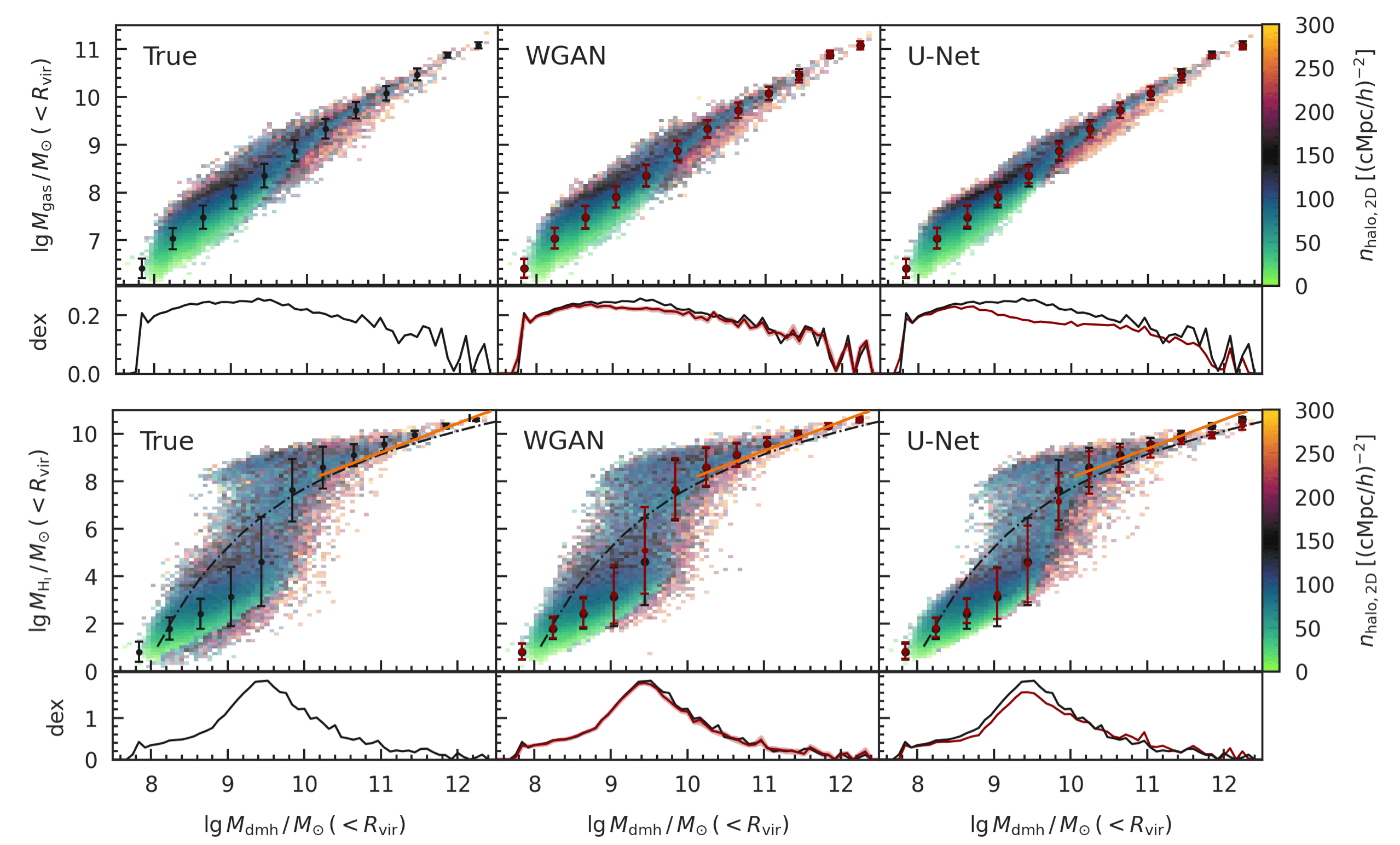}
    \caption{Summary figure of the halo analysis showing the relation between the projected gas (upper panel), $\HI$ (lower panel) and dark matter mass within one projected virial radius. Bins are color-coded by the median halo number density as a proxy for the halo environment where a higher transparency indicates fewer data points. We also show quantiles (16, 50 and 84) computed on the binned data. Also shown is the $\HI$ abundance matching result at $z=2$ from \protect\cite{Padmanabhan2016b} in orange and the fit from \protect\cite{Villaescusa2018} as the black dash-dotted line. The curves in the lower panels indicate the scatter across the range of halo masses. Red lines with shaded area denote the (16, 50 and 84) quantiles for the 128 predicted boxes with WGAN whereas only the median is shown in the U-Net case. Black symbols denote true data points, which are also shown in the WGAN and U-Net panels for better visual comparison. Red symbols denote the corresponding predictions.}
    \label{fig:1024_haloes}
\end{figure*}

We show the results of this analysis in Fig.~\ref{fig:1024_haloes}. To quantify the intrinsic scatter in the dataset we bin the dark matter masses and compute the median gas masses for each bin, whereas the lower panel indicates the true and predicted amount of intrinsic scatter. Interestingly, the deterministic U-Net shows non-zero scatter across all halo masses and the WGAN prediction is in even better agreement with the simulation. Furthermore, Fig.~\ref{fig:1024_haloes} exhibits a very interesting aspect. The exact amount of gas does not just simply depend on the halo mass alone, but also on the dark matter environment. Dark matter haloes of the same mass but at different locations in the cosmic web can contain varying amounts of gas. For a fixed halo mass bin the scatter in the gas and $\HI$ counterparts can be large depending on the cosmic environmental density as well as the surrounding gas reservoir on Mpc-scale. Since the networks have access to this large scale information, the predictions can account for the dark matter environment. To emphasize this aspect, mass bins are color-coded according to their median 2D number density in Fig.~\ref{fig:1024_haloes}, which is computed by counting the number of neighbouring haloes within a $r_0=1\,h^{-1}\text{Mpc}$ disk, centered at the halo of interest, i.e.
\begin{equation}
    n_\mathrm{halo, 2D} = \frac{N(<r_0)}{\pi r_0^2}.
\end{equation}
Furthermore, the halo based analysis suggests that at fixed redshift the dark matter cosmic web contains sufficient information to reconstruct the gas mass distribution \citep[see e.g.][]{Kraljic2019}. Apparently, the required information from the halo growth history is encoded in the environment such that including information from previous redshifts is not strictly necessary.\\

The \fb simulation shows large scatter in $M_{\HI}$ for haloes below $10^{11}\,\Msun$. We find a scatter of order 0.5 dex, with an extreme case of 2 dex around $10^{10}\, \Msun$. The halo mass to $\HI$ mass relation has previously been modelled with abundance matching (AM) techniques \citep[see e.g.][]{Papastergis2013, Padmanabhan2016a, Padmanabhan2016b, Spina2021}. Typically, these models target halo masses above $\sim 10^{10} \, \Msun$ as they are constrained from observational data. 
The large scatter for low mass haloes indicates that simple AM models might break down here as more information about the dark matter halo is necessary to accurately predict the contained $\HI$ mass. We show this behaviour in Fig.~\ref{fig:1024_haloes} where we compare our results to the AM predictions of \cite{Padmanabhan2016b} at $z=2$. For large halo masses, our sample shows a small scatter and follows the theoretical linear relation very closely. 
However for halo masses below $10^{10}\,\Msun$ a linear relation is a poor description of the data as there are subgroups that tend to populate distinct places in the mass plane. We argue that our neural network approach is better suited to model the mapping in these regimes because it accounts for the higher complexity of the relation. At fixed halo mass the $\HI$ mass correlates with the environmental density. Fig. \ref{fig:1024_haloes} shows that the $\HI$ content in haloes with different environments (colors indicating different halo number densities) as well as the predicted scatter, are almost perfectly reconstructed by the neural network indicating that the environment is necessary for learning the mean and scatter of the relation.
Overall, our WGAN model makes more accurate predictions than the U-Net, which underestimates the scatter for halo masses $10^{8}-10^{10}\,M_{\odot}$. relation.
We therefore conclude that the WGAN offers a valuable approach to model the dark matter to $\HI$ mass relation across the entire range probed by the training data and especially in the low mass end where environmental effects become noticeably more important.

\subsubsection*{Application to B100}\label{sec:application_to_b100}

\begin{figure*}
    \includegraphics[width=\textwidth]{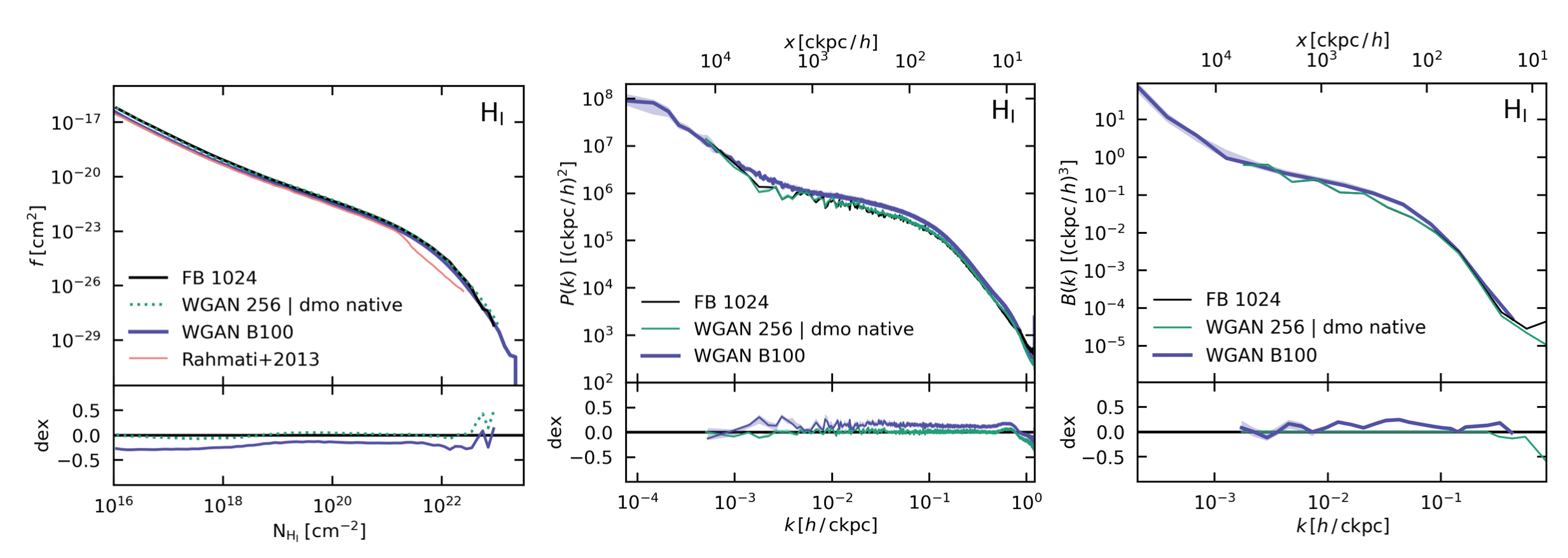}
    \caption{Here we show the CDDF, power spectrum and bispectrum of the B100 prediction. We also overplot the statistics from the \fb simulation (\textit{HR}, labelled as FB1024) and the WGAN prediction on the native dark-matter-only \textit{MR} FIREbox (labelled as WGAN 256 | dmo native) to emphasize how the WGAN model can be used on large dark matter simulations to extend the summary statistics, e.g. probing lower $\HI$ column densities. Shaded violet areas denote the 10\% error limit, representing the internal scatter of the WGAN predictions.}
    \label{fig:b100}
\end{figure*}

\begin{figure*}
    \includegraphics[width=\textwidth]{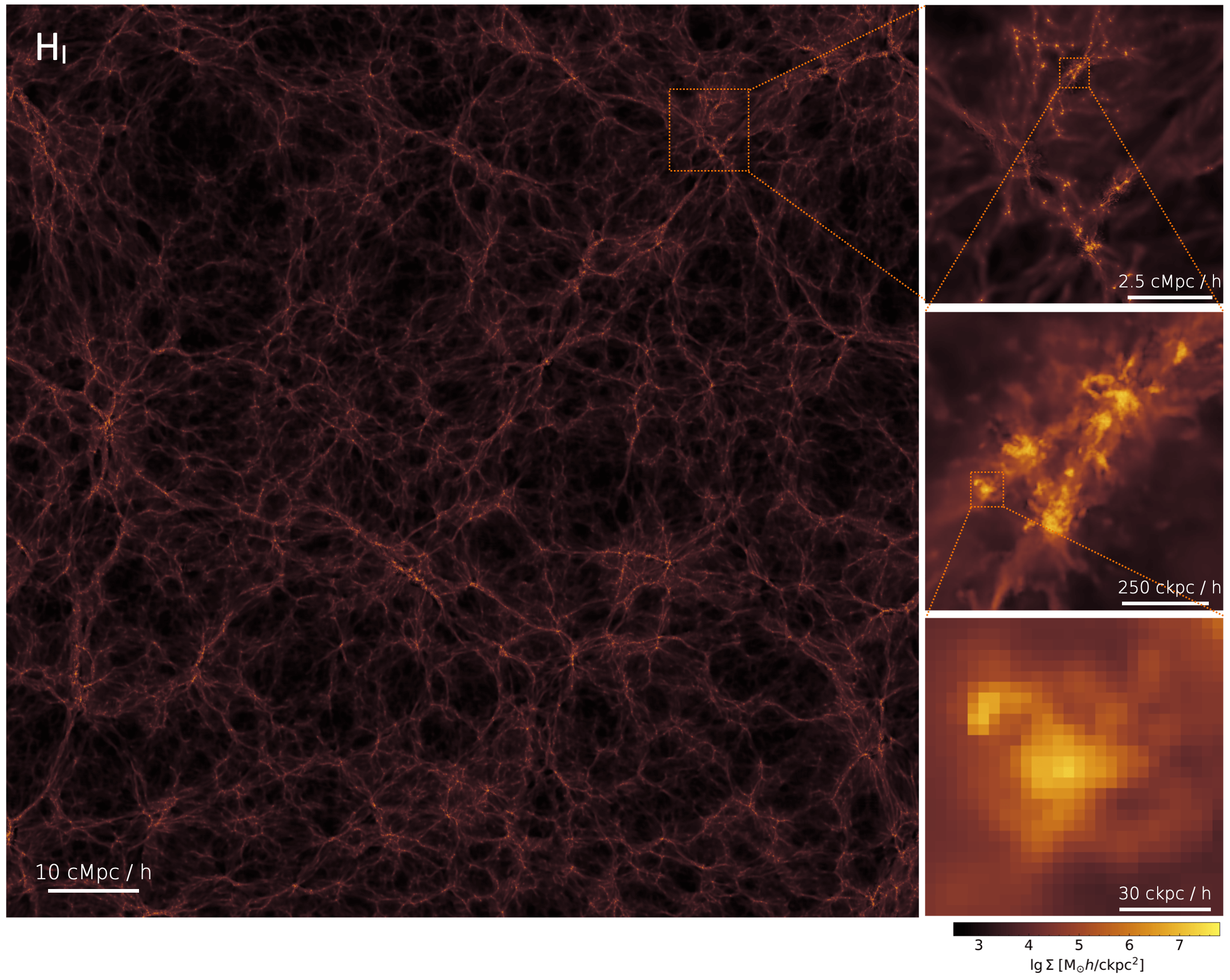}
    \caption{Illustration of an individual slice of $1.5\,\text{cMpc}\,h^{-1}$ depth and $100\,\text{cMpc}\,h^{-1}$ in size of the emulated $\HI$ field using our fiducial WGAN $\HI$ model. The full image shown is reduced in pixel resolution compared to the original version (due to filesize limitations) that consists of $27307^{2}$ pixels, where one pixel resolves $\sim 3.6\,\text{ckpc}\,h^{-1}$. The full map can be found at the official \href{https://www.github.com/maurbe/ember}{github} repository.}
    \label{fig:b100_zoom}
\end{figure*}

Fig.~\ref{fig:HI_summary} shows that the WGAN prediction on the power and bispectrum is very similar across the different input fields, despite the model being trained on \texttt{dmh ds} as input. This behaviour is crucial when applying the model to enrich dark matter only simulations with larger box sizes. To demonstrate the pipeline, we predict the $\HI$ maps for the B100 simulation, which has very similar mass resolution properties as the \textit{MR} training set (see Table \ref{tab:1} for details). Since the generator network is fully convolutional, we can predict entire $\HI$ slices simultaneously, eliminating the problem of edge effects. However, due to the large memory consumption, this operation is currently only possible on CPUs. The prediction of one $\HI$ mass map takes $\sim$1 hour. We show an example slice of the projected box in Fig.~\ref{fig:b100_zoom}. Fig.~\ref{fig:b100} displays the summary statistics of the B100 predictions, which nicely extend and match the \textit{HR} \fb and \textit{MR} WGAN prediction on large and small scales.

The CDDF is a measure of the number of $\HI$ absorbers per unit column density. Fig.~\ref{fig:b100} demonstrates how the network extrapolates the CDDF for B100 with respect to \fb when probing a larger volume. In all our experiments we observe a steepening of the CDDF beyond $\log\,\text{N}_{\HI} \geq 21$, which then linearly extends towards higher density systems. This result is in good agreement with related work in \cite{Rahmati2014} up to $10^{21}\text{cm}^{-2}$. \cite{Rahmati2013} showed that in order to have a converged CDDF up to $10^{22}\text{cm}^{-2}$ one requires a box size of at least 50 Mpc/h. Furthermore, when identifying sightlines with grid cells, a coarser grid will introduce an artificial smoothing, effectively shifting the CDDF to lower number densities. This is an additional advantage of our method, as it allows to compute column densities over high resolution grids for large boxes. The upsampling capabilities in our approach are especially interesting for creating mock observations for applications where a broad range of resolved halo masses is necessary. One such application is Intensity Mapping (IM) \citep[][]{Kovetz2019, Padmanabhan2019}, where a significant contribution to the $\HI$ signal is expected to come from galaxies residing in lower mass haloes of $\approx 10^9 h^{-1} \Msun$ \citep[][]{Villaescusa2018, Cunnington2018}. Simultaneously, realistic mock observations also require simulations of large box sizes -- hundreds of Mpc to Gpc -- to supply the wide and deep light-cones relevant to IM studies. The EMBER approach attempts to bridge these two extreme regimes by construction.

We also show the power and bispectrum of the B100 prediction compared to the \fb simulation and the WGAN prediction on the \textit{MR} \texttt{dmo native}. The statistics of the power and bispectrum are in good agreement between the three different predictions but the B100 statistics slightly differ on intermediate scales. On one hand, cosmic variance might be one possible explanation of this behaviour as we only have one \fb realization to compare against. On the other hand, the effects from large scale modes might be important as well. \cite{vanDaalen2019} found that the most massive haloes have an effect on the power spectrum as they contribute to the power on scales $k \geq 10^{-2} \, h \, \text{ckpc}^{-1}$. This aspect was pointed out before by \cite{Chisari2018}, who compared power spectra from sub-volumes drawn from their fiducial $100 \, h^{-1} \, \text{cMpc}$ simulation. They found significant variation between samples, depending on whether a massive object was present in a given sub-volume or not. For the comparison of the B100 with \fb in Fig.~\ref{fig:b100}, these results indicate that the slight difference in power is a combination of cosmic variance and the more massive haloes affecting intermediate scales.

Baryons modify the full density field and thus affect the entire hierarchy of higher order statistics beyond the power spectrum \citep[][]{Arico2020}. \cite{Foreman2019} showed that baryonic effects on the bispectrum of hydrodynamical simulations carry additional information with respect to the power spectrum. As showed in \cite{Foreman2019} and \cite{Arico2020}, massive haloes contribute to baryonic effects more in the bispectrum than in the power spectrum. Consequently, the bispectrum measured in relatively small boxes is different at small scales with respect to larger boxes. This effect is similar to the one for the power spectrum. To mitigate this issue we included four zoom in simulations of massive haloes such that the training data represented the modes of individual massive haloes.


\section{Summary and Conclusions}\label{sec:conclusion}
We have presented EMBER, a novel deep-learning-based framework to emulate baryonic maps, specifically for gas and atomic hydrogen, from dark matter data alone. Our training data is based on hydrodynamical simulations run with the FIRE-2 physics model at high numerical resolution ($m_{\rm b}=3-6\times{}10^4$ $M_\odot$). This simulation suite includes a $(15\,\text{cMpc}/h)^3$ cosmological volume simulation (\fb) and several zoom-in simulations of massive haloes. We emphasize that combining small cosmological boxes with high-resolution zoom-in simulations is a crucial ingredient for applying our model to predict baryon fields for dark matter simulations of larger box sizes. This methodology ensures that the entire range of scales is represented in the training set.

By applying EMBER to test data, we found that it is able to reproduce important physical statistics such as power and bispectra as well as the $\HI$ CDDF. In particular, \fb predicts that the relationship between halo mass and $\HI$ galaxy mass is environmentally dependent and breaks dramatically with significant scatter around a characteristic halo mass scale $\sim 5\cdot 10^{9}\, M_\odot$.  Such a trend would be difficult to capture in a traditional halo-mass-based approach, while EMBER is able to reproduce the relationship with remarkable accuracy using only the dark matter field (see Figure \ref{fig:1024_haloes}). Furthermore, we showed that EMBER is capable of emulating high resolution baryon information from low resolution dark matter inputs through upsampling techniques. This is an extremely attractive property of our approach, as it allows to populate large, but low resolution, dark matter simulations with baryon fields at the high resolution level of the training data. In the following we summarize our main findings in more detail.

\begin{itemize}
    \item We showed that a stochastic WGAN architecture is able to capture and learn the feature distribution on very small scales better than a U-Net model. We conclude that the additional variance that the WGAN offers is necessary to accurately model the mapping from dark matter to baryons especially on scales $\leq 100$ ckpc/$h$.
    \item In particular, we found that the $\HI$ U-Net struggles when comparing the total mass in the box. \cite{Wadekar2020} used a second network trained just on the high density pixels to mitigate this problem. However, we found that the WGANs do not suffer from this problem and adapt well to the different target fields.
    \item The WGAN models are able to reproduce the power spectra on the corresponding gas targets with $\sim10\%$ accuracy down to $\sim10 \, \text{ckpc}/h$. Furthermore, we found very good agreement on the pixel PDF and the CDDF. The bispectra of the predictions are within 20\% accuracy indicating that the adversarial networks are capable of capturing even higher order moments in the data set such as the filamentary structures between haloes.
    \item We conducted a halo-based analysis to compare the true and predicted gas masses within one virial radius of parent dark matter haloes.
    The network predictions agree very well with \fb and the analytical AM relation from \cite{Padmanabhan2016b}, indicating that the network has learned to retrieve features in the surrounding of the dark matter haloes to determine the contained gas mass. This is a big advantage of EMBER as it defines a halo-free method \citep[compare e.g. with][]{Lovell2021} that is sensitive to the dark matter environment on a large range of scales and can thus be used to extend the result of current AM techniques down to very low halo masses. Furthermore, the reproduced scatter in the gas mass at fixed halo masses is very close to the intrinsic scatter in the dataset despite the fact that the model inputs have no dynamical information. We will investigate this direction in the future to better understand the main drivers that determine the gas masses and scatter in haloes and how it depends on the underlying dark matter morphology.
    \item We investigated the case of predicting the target fields from maps that are derived from lower resolution dark matter simulations and compared them with the high resolution targets. From the analysis we conclude that the networks were still able to make accurate predictions in terms of the summary statistics when reducing the input resolution by two levels to $256^{3}$, i.e. a factor 64 in mass.
    \item We conducted the same analysis for the extreme upsampling cases by reducing the dark matter resolution by four levels $(64^3)$ and six levels $(16^3)$ and found reasonably accurate predictions (typically $\sim 20 \%$ error) on the gas and $\HI$ power spectrum and bispectrum down to scales $\sim 50 \, \text{ckpc}/h$. However, the extreme upscaling from $16^{3}$ to $1024^{3}$ results in errors far beyond 10\% when predicting the PDF, CDDF and the total mass in the box.
    \item We quantitatively illustrated the application of how the WGAN models can be used to predict gas maps for larger cosmological volumes. More specifically, we applied the WGAN $\HI$ network trained on our hydrodynamical simulation suite to emulate $\HI$ mass maps for a $100\, \text{cMpc}/h$ dark matter simulation. On small scales, the predicted gas and $\HI$ power spectrum and CDDF agree well with those in our simulation suite, while on large scales they are in good agreement with the results by \cite{Rahmati2014}.
\end{itemize}
Overall, our fiducial WGAN approach shows very good agreement for all the tested metrics and has excellent upsampling capabilities when presented with lower resolution information.

\subsection*{Future work}
In the current implementation, EMBER regresses only one target field from one single input channel. One promising extension would be to add more dark matter inputs, e.g. the velocity field or the dispersion thereof to include dynamical information. Using the velocity dispersion as an additional feature might allow to identify merging systems and subsequently information on the recent assembly history and the gas content.

Furthermore, the network architecture is easily extended to predict additional baryon fields such as temperature, pressure or stellar densities. Since the noise input in the WGAN models is responsible for generating the small scale features, training on multiple target fields simultaneously ensures that predictions of different baryon fields are coherent when sampling. This is an important advantage as opposed to training multiple networks that each predict one individual field.

Since our networks were trained on data from one redshift, this approach currently works for this specific epoch. In principle one can train the same network architectures on any specific redshift. Furthermore, it would be interesting to allow redshift interpolation by training on all redshift slices. We will investigate this path in future work.

Currently, our methodology operates on 2-dimensional projection maps that aggregate information over a slab thickness of $1.5\,h^{-1}\text{Mpc}$. This approach effectively produces tomographic maps of the underlying 3-dimensional volume. 
Tomographic approaches have recently been used to reconstruct from observations 3-dimensional density maps to compute Lyman-alpha flux profiles \citep[e.g.][]{Lee2018, Newman2020, Li2021}. 
Our approach is well matched with current line-of-sight resolution limits of spectrographs such as e.g. MUSE \citep[][]{MUSE2010}, which are in fact of order $\sim$$\text{Mpc}$ \citep[see e.g.][and references therein]{Ravoux2020}.

We also conducted a halo-based analysis to understand to what extent different dark matter environments influence the prediction of the gas masses contained in the haloes. In future work, we will investigate a framework to quantify the importance of the dark matter environment for the predictions, since this is the major advantage that our methodology offers compared to halo-based models. Finally, the methodology of EMBER presented in this work promises an exciting pathway for fast and accurate enrichment of dark matter only simulations with high resolution baryon information.

\section*{Acknowledgements}
MB thanks Tomasz Kacprzak for fruitful discussions and Darren Reed for technical support for the GPU training. Furthermore the authors want to thank Claude-André Faucher-Giguère, Hugues Lascombes, Romain Teyssier and Aurel Schneider for helpful comments and insights that helped improve this work. RF acknowledges financial support from the Swiss National Science Foundation (grant no 157591 and 194814). DAA acknowledges support by NSF grant AST-2009687 and by the Flatiron Institute, which is supported by the Simons Foundation. MBK acknowledges support from NSF CAREER award AST-1752913, NSF grant AST-1910346, NASA grant NNX17AG29G, and HST-AR-15006, HST-AR-15809, HST-GO-15658, HST-GO-15901, HST-GO-15902, HST-AR-16159, and HST-GO-16226 from the Space Telescope Science Institute, which is operated by AURA, Inc., under NASA contract NAS5-26555. We acknowledge PRACE for awarding us access to MareNostrum at the Barcelona Supercomputing Center (BSC), Spain. This research was partly carried out via the Frontera computing project at the Texas Advanced Computing Center. Frontera is made possible by National Science Foundation award OAC-1818253. This work was supported in part by a grant from the Swiss National Supercomputing Centre (CSCS) under project IDs s697 and s698. We acknowledge access to Piz Daint at the Swiss National Supercomputing Centre, Switzerland under the University of Zurich’s share with the project ID uzh18. This work made use of infrastructure services provided by S3IT (www.s3it.uzh.ch), the Service and Support for Science IT team at the University of Zurich. Finally, we thank the referee Simeon Bird for his valuable input that helped improve this work.

\section*{Data Availability Statement}
We provide the source code, networks and maps at the official repository: \url{https://github.com/maurbe/ember}.
\newpage



\bibliographystyle{mnras}
\bibliography{document}


\appendix
\section{CDDF}\label{app:A}
In this section we discuss important aspects regarding the computation of the CDDF for simulation data with finite box size $L$. The absorption length $X(z)$ depends on the redshift $z$ via the following integral expression \citep{Peebles1969, Nagamine2004}
\begin{equation}
    X(z) = \int_{0}^{z} (1+x)^{2}\frac{H_0}{H(x)} \diff x.
\end{equation}
The comoving distance $L$ to redshift $z$ is determined by the cosmology through
\begin{equation}
    L = c \int_{0}^{z} \frac{\diff x}{H(x)} \quad \text{with} \quad \diff L=\frac{c}{H(z)} \diff z.
\end{equation}
The differential absorption length $\Delta X$ is then computed as following
\begin{align}
\begin{split}
    \frac{\diff X}{\diff L} &= \frac{\diff}{\diff L} \int_{0}^{z} (1+x)^{2} \frac{H_0}{H(x)} \diff x\\
    &= \frac{H(z)}{c} \frac{\diff}{\diff z} \int_{0}^{z} (1+x)^{2} \frac{H_0}{H(x)} \diff x
    = \frac{H_0}{c} (1+z)^{2}.
\end{split}
\end{align}

\begin{figure}
    \includegraphics[width=\columnwidth]{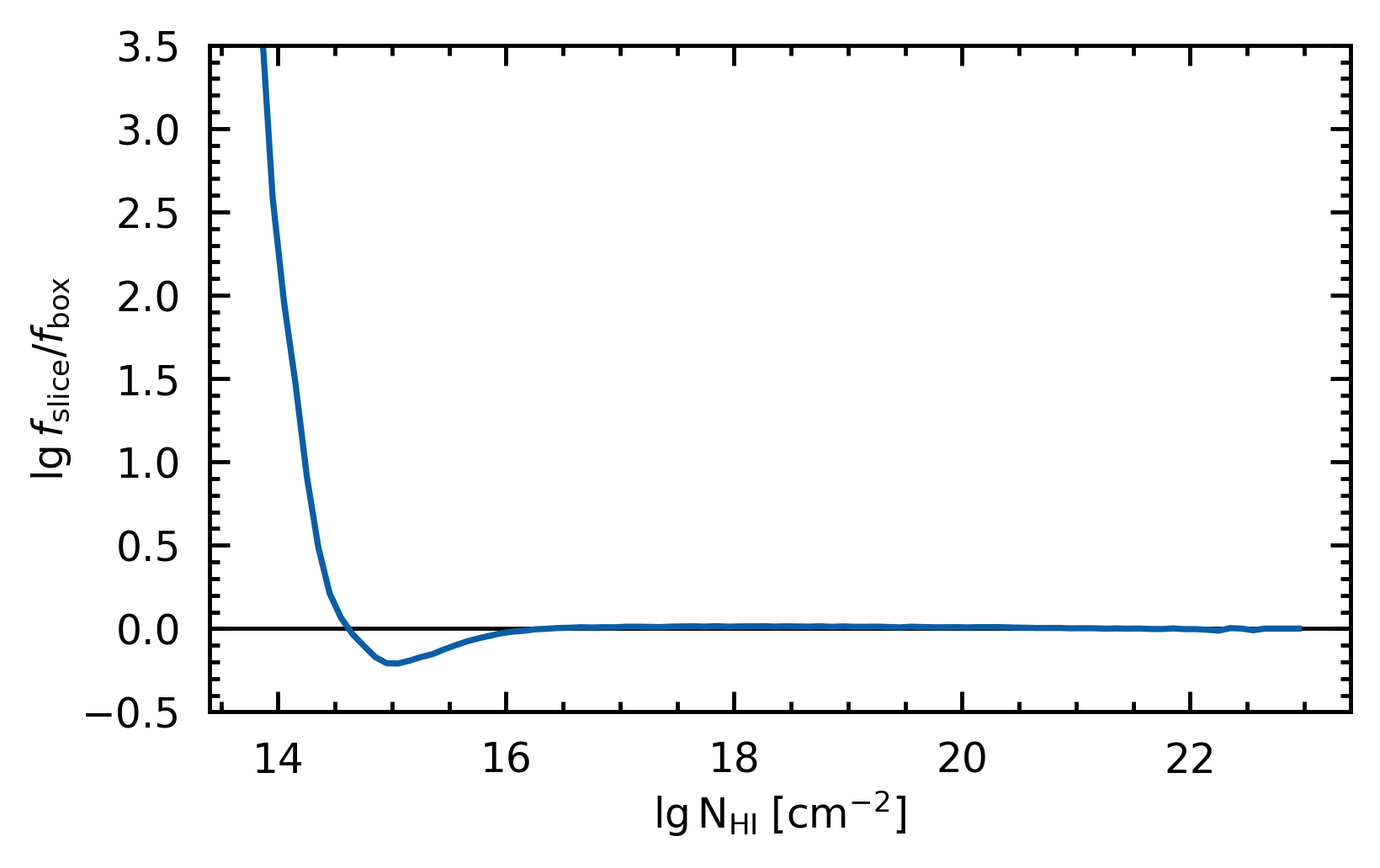}
    \caption{We show the ratio of the $\HI$ CDDF in the \fb simulation when computing it by accumulating pixels over individual slabs compared to the CDDF of the projection of the total box. The comparison indicates that for systems above $\text{N}_{\text{H}_{\text{I}}} > 10^{16}\mathrm{cm}^{-2}$ the statistics are identical.}
    \label{fig:A1}
\end{figure}

The CDDF is a pixel-based statistic computed over 2-dimensional projection maps. The exact shape of the CDDF therefore depends on the projection width of individual slabs, because individual systems can in principle overlap. We check that the CDDF has converged for our use-case by computing it once for the $\HI$ maps projected over the entire box ($f_{\text{total box}}$) and once for the CDDF accumulated over all slabs $f_{\text{slabs}}$ (shown in Fig. \ref{fig:A1}). The comparison study clearly shows that the CDDF is identical for the two approaches above $N_{\text{H}_{\text{I}}}>10^{16}\text{cm}^{-2}$. Hence, systems with $\lg N_{\HI} > 16$ rarely overlap in the box. Since we are not interested in lower column density systems in this study, we conclude that either approach is applicable for our prediction pipeline.

\section{Halo based analysis}\label{app:B}
\begin{figure}
    \includegraphics[width=\columnwidth]{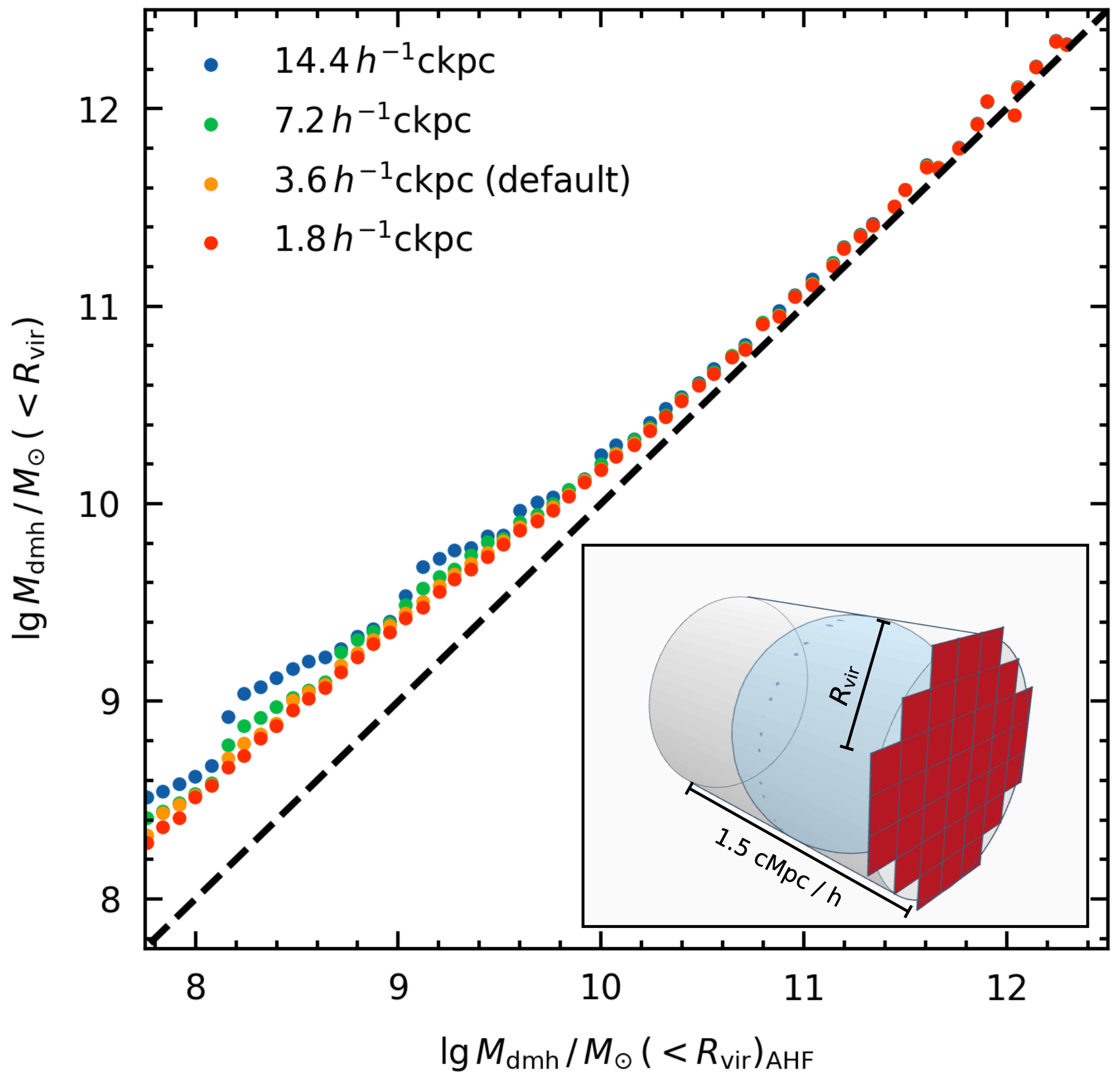}
    \caption{This figure shows a comparison between the median dark matter masses within one virial radius and the projected halo masses over one slab thickness of $1.5\,h^{-1}\text{cMpc}$. The inset figure illustrates our approach. The mass within one virial radius as computed by AHF is shown in blue. The cylinder over which the masses are projected in our approach is shown in grey. Since our approach operates upon grids, we also show an example of a pixelized halo mask in red. The x-axis corresponds to the true AHF mass in blue while the y-axis is the projected mass shown in red. We also show this relation for different pixel resolutions indicated in the legend. Note that $3.6\,h^{-1}\text{ckpc}$ corresponds to the resolution that is used throughout this work.}
    \label{fig:halo_check}
\end{figure}

In Fig.~\ref{fig:halo_check} we show a more detailed comparison between dark matter halo masses computed from AHF and the projected masses that we used for our halo based approach. The figure compares these two masses and shows how the offset changes with resepect to the grid resolution and mass scale. Clearly, the projected halo masses are systematically larger than the AHF masses. This trend becomes stronger for smaller halo masses (and thus smaller virial radii). The explanation of this effect is two-fold. First, the grid resolution plays an important role as higher resolution masks can approximate the exact virial radius better and thus get a better estimate of the contained mass. We show this analysis for different pixel resolutions and conclude that at our resolution level of $3.6\,h^{-1}\text{ckpc}$ this relation is converged. Second, projecting the mass over an entire slab will overestimate the masses within individual haloes. This effect becomes stronger for smaller haloes as the contamination becomes higher. The largest differences are of order ~0.5 dex and for halo masses $\log (M_{\text{dmh}} /\Msun)(<R_{\text{vir}}) > 10.5$ the two masses are almost identical.

\begin{figure*}
    \includegraphics[width=\textwidth]{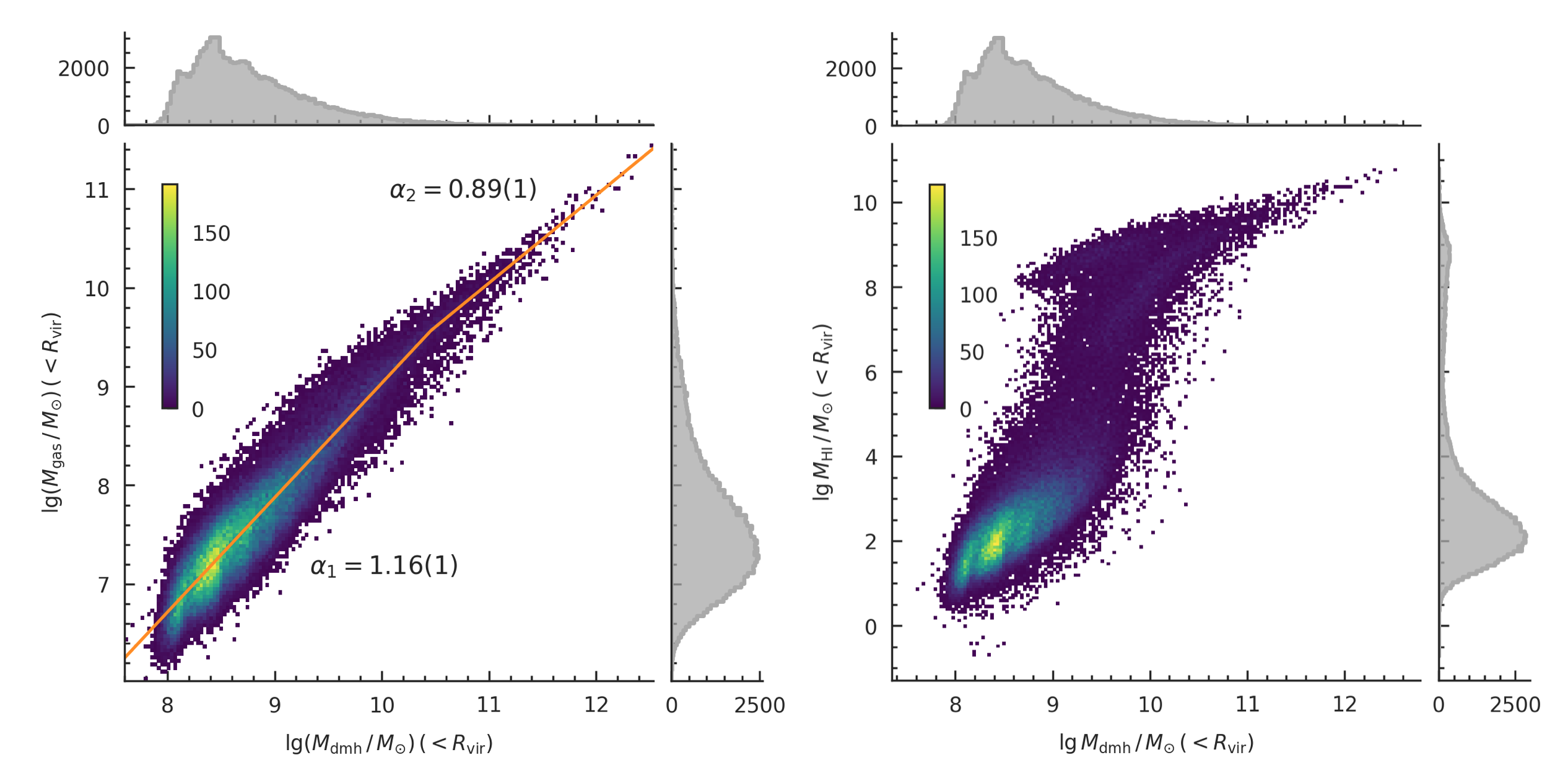}
    \caption{Joint distributions showing the projected dark matter and gas masses in individual haloes corresponding to Fig.~\ref{fig:1024_haloes}. We provide the underlying data at the official github repository \url{https://github.com/maurbe/ember}.}
    \label{fig:haloes_joint}
\end{figure*}

In Fig.~\ref{fig:haloes_joint} we show the halo counts in the \fb simulation for individual mass bins as described in section \ref{sec:halo_analysis}. For the gas the distribution is fit with a broken power law with breaking point at $\log(M_{\text{dmh}}/M_{\odot})_{\text{break}}=10.46(2)$ and slope parameters $\alpha_1=1.16(1)$ and $\alpha_2=0.89(1)$.

\bsp	
\label{lastpage}

\end{document}